\documentclass[11pt]{article}
\usepackage{graphicx}
\usepackage{amssymb,amsmath,amsfonts}

\def\Re{{\cal R \mskip-4mu \lower.1ex \hbox{\it e}\,}}
\def\Im{{\cal I \mskip-5mu \lower.1ex \hbox{\it m}\,}}
\def\ie{{\it i.e.}}
\def\eg{{\it e.g.}}
\def\etc{{\it etc}}

\def\sub#1{_{\lower.25ex\hbox{$\scriptstyle#1$}}}
\def\tev{\,{\ifmmode\mathrm {TeV}\else TeV\fi}}
\def\gev{\,{\ifmmode\mathrm {GeV}\else GeV\fi}}
\def\mev{\,{\ifmmode\mathrm {MeV}\else MeV\fi}}
\def\mpl{\ifmmode M_{pl}\else $M_{pl}$\fi}
\def\mpl{\ifmmode \overline M_{Pl}\else $\bar M_{Pl}$\fi}
\def\to{\rightarrow}

\def\subw{_{\rm w}}
\def\mh{\ifmmode m\sbl H \else $m\sbl H$\fi}
\def\mch{\ifmmode m_{H^\pm} \else $m_{H^\pm}$\fi}
\def\mt{\ifmmode m_t\else $m_t$\fi}
\def\mc{\ifmmode m_c\else $m_c$\fi}
\def\mz{\ifmmode M_Z\else $M_Z$\fi}
\def\mw{\ifmmode M_W\else $M_W$\fi}
\def\mws{\ifmmode M_W^2 \else $M_W^2$\fi}
\def\mhs{\ifmmode m_H^2 \else $m_H^2$\fi}   
\def\mzs{\ifmmode M_Z^2 \else $M_Z^2$\fi}
\def\mts{\ifmmode m_t^2 \else $m_t^2$\fi}
\def\mcs{\ifmmode m_c^2 \else $m_c^2$\fi}
\def\mchs{\ifmmode m_{H^\pm}^2 \else $m_{H^\pm}^2$\fi}
\def\ztwo{\ifmmode Z_2\else $Z_2$\fi}
\def\zone{\ifmmode Z_1\else $Z_1$\fi}
\def\mtwo{\ifmmode M_2\else $M_2$\fi}
\def\mone{\ifmmode M_1\else $M_1$\fi}
\def\tb{\ifmmode \tan\beta \else $\tan\beta$\fi}
\def\xw{\ifmmode x\subw\else $x\subw$\fi}
\def\ch{\ifmmode H^\pm \else $H^\pm$\fi}
\def\lum{\ifmmode {\cal L}\else ${\cal L}$\fi}
\def\inpb{\,{\ifmmode {\mathrm {pb}}^{-1}\else ${\mathrm {pb}}^{-1}$\fi}}
\def\infb{\,{\ifmmode {\mathrm {fb}}^{-1}\else ${\mathrm {fb}}^{-1}$\fi}}
\def\epem{\ifmmode e^+e^-\else $e^+e^-$\fi}
\def\ppb{\ifmmode \bar pp\else $\bar pp$\fi}
\def\bsg{\ifmmode B\to X_s\gamma\else $B\to X_s\gamma$\fi}
\def\bsll{\ifmmode B\to X_s\ell^+\ell^-\else $B\to X_s\ell^+\ell^-$\fi}
\def\bstt{\ifmmode B\to X_s\tau^+\tau^-\else $B\to X_s\tau^+\tau^-$\fi}
\def\lamt{\ifmmode \tilde\lambda\else $\tilde\lambda$\fi}
\def\shat{\ifmmode \hat s\else $\hat s$\fi}
\def\that{\ifmmode \hat t\else $\hat t$\fi}
\def\uhat{\ifmmode \hat u\else $\hat u$\fi}

\newskip\zatskip \zatskip=0pt plus0pt minus0pt
\def\matth{\mathsurround=0pt}
\def\lsim{\mathrel{\mathpalette\atversim<}}
\def\gsim{\mathrel{\mathpalette\atversim>}}
\def\atversim#1#2{\lower0.7ex\vbox{\baselineskip\zatskip\lineskip\zatskip
  \lineskiplimit 0pt\ialign{$\matth#1\hfil##\hfil$\crcr#2\crcr\sim\crcr}}}

\def\grtsim{\,\,\rlap{\raise 3pt\hbox{$>$}}{\lower 3pt\hbox{$\sim$}}\,\,}
\def\lsim{\,\,\rlap{\raise 3pt\hbox{$<$}}{\lower 3pt\hbox{$\sim$}}\,\,}

\renewcommand{\thefootnote}{\fnsymbol{footnote}}

\begin{document} \begin{titlepage}
\rightline{\vbox{\halign{&#\hfil\cr
&SLAC-PUB-11864\cr
}}}
\begin{center}
\thispagestyle{empty} \flushbottom { {
\Large\bf Noncommutative Inspired Black Holes in Extra Dimensions
\footnote{Work supported in part
by the Department of Energy, Contract DE-AC02-76SF00515}
\footnote{e-mail:
$^a$rizzo@slac.stanford.edu}}}
\medskip
\end{center}

\centerline{Thomas G. Rizzo$^{a}$}
\vspace{8pt} 
\centerline{\it Stanford Linear
Accelerator Center, 2575 Sand Hill Rd., Menlo Park, CA, 94025}

\vspace*{0.3cm}

\begin{abstract}
In a recent string theory motivated paper, Nicolini, Smailagic and Spallucci 
(NSS) presented an interesting model for a noncommutative inspired,  
Schwarzschild-like black hole solution in 4-dimensions. The essential 
effect of having noncommutative co-ordinates in this approach is to smear out 
matter distributions on a scale associated with the turn-on of noncommutativity 
which was taken to be near the 4-d Planck mass. In particular, NSS took 
this smearing to be essentially Gaussian. This energy scale is sufficiently 
large that in 4-d such effects may remain invisible indefinitely. Extra 
dimensional models which attempt to address the gauge hierarchy problem, 
however, allow for the possibility that the effective fundamental scale may 
not be far from $\sim$ 1 TeV, an energy regime that will soon be probed by 
experiments at both the LHC and ILC. In this paper we generalize the NSS 
model to the case where flat, toroidally compactified extra dimensions are 
accessible at the Terascale and examine the resulting modifications in black  
hole properties due to the existence of noncommutativity. We show that while 
many of the noncommutativity-induced black hole features found in 4-d by NSS 
persist, in some cases there can be significant modifications due the 
presence of extra dimensions. We also demonstrate that the essential 
features of this approach are not particularly sensitive to the Gaussian 
nature of the smearing employed by NSS. 
\end{abstract}



\renewcommand{\thefootnote}{\arabic{footnote}} \end{titlepage} 

%
%
%

\section{Introduction and Background}

The theoretical effort that has gone into understanding the full details of string/M theory has inspired a 
number of ideas which, on their own, have had a significant impact on particle physics model building and phenomenology. 
One of the more recent developments of this kind has been the resurgence{\cite {old}} of interest in noncommutative (NC) 
quantum field theories{\cite {NCQFT}} and, in particular, the question of how a NC version of the Standard Model (SM) 
may be constructed{\cite {NCSM}} and probed experimentally{\cite {NCPHENO}}. 

The essential idea behind NC constructions is that the commutator of two spacetime coordinates, now thought of as 
operators, is no longer zero. 
In its simplest form, for a space with an arbitrary number of dimensions, $D$, this is can be written explicitly as  
\begin{equation}
[x_A,x_B]=i\theta_{AB}=i{c_{AB}\over {\Lambda_{NC}^2}}\,,
\end{equation}
where $\Lambda_{NC}$ is the mass scale associated with NC 
and $c_{AB}$ is normally taken to be a {\it frame-independent}, dimensionless anti-symmetric matrix with constant,  
real, typically $O(1)$ entries; it is {\it not} a tensor. Here we assume that vastly different NC scales do not exist 
depending upon the values 
of $A,B$. {\footnote {In what follows, upper case Roman letters will label all $D$ dimensions while Greek (lower case 
Roman) letters will cover the range 0-3 (5 to $n+4$).}}  In a general string theory context one might imagine that 
$\Lambda_{NC}$ would naturally not be far 
from the 4-d Planck scale, $\mpl$, and that the $c_{AB}$ are generated due to the  
presence of background `electric' or `magnetic' type fields. Most of the phenomenological studies of NC 
models{\cite {NCPHENO}} have assumed that we live in 4-d and that $\Lambda_{NC} \sim$ 1-10 TeV so that we have access to 
this scale at, \eg, the LHC or ILC. However, if the NC scale $\Lambda_{NC}$ is indeed of order $\mpl$, then probing NC 
physics directly may prove difficult in the near term. 

One way of possibly observing NC is its effects on the properties of black holes (BH). In order to analyze this problem 
at a truly fundamental level one would need to successfully construct the NC equivalent of General Relativity. 
Attempts along these lines have been made in the literature{\cite {NCGR}} but no complete and 
fully compelling theory of this type 
yet exists. Recently, Nicolini, Smailagic and Spallucci (NSS) {\cite {Nicolini:2005vd}} have considered a physically 
motivated and tractable model of the possible NC modifications to Schwarzschild BH 
solutions. The essential ideas of this picture are: ($i$) General Relativity in its usual commutative form as 
described by the Einstein-Hilbert action remains applicable. This seems justifiable, at least to a good approximation, if 
NC effects can be treated perturbatively. The authors in Ref.{\cite {NCGR}} have indeed shown that the leading NC corrections 
to the form of the Einstein-Hilbert action are at least second order in the $\theta_{AB}$ parameters. ($ii$) NC leads to 
a `smearing' of matter distributions on length scales of order 
$\sim \Lambda_{NC}^{-1}$. Thus the usual `$\delta$-function' matter source of the conventional Schwarzschild solution is 
replaced by a centrally peaked, spherically symmetric (and time-independent) mass distribution which has a size of order 
$\sim \Lambda_{NC}^{-1}$. This, too, seems justifiable based on the results presented in Ref.{\cite {NCGR}}, which  
note that matter actions from which the stress-energy tensors are derived are modified at leading order in the $\theta_{AB}$ 
parameters.  Based on earlier work{\cite {Gauss}}, NSS took this smeared distribution to be in the form of a spherical 3-d 
Gaussian in 4-d whose size, due to the spherical symmetry, was set by a single parameter, $\theta$, indicative of the NC 
scale. Though such a picture leads to many interesting properties for the resulting BH (to be elaborated on below), 
since $\mpl \sim \Lambda_{NC}$ was assumed, as would be natural in 4-d, such BH are not immediately 
accessible to experiment or to direct observation. 

Another interesting prediction of string theory is that several extra dimensions must exist. 
However, extra spatial dimensions, in models with an effective fundamental scale $M_*$ now in the TeV range, have 
been discussed as possible solutions  
to the hierarchy problem{\cite {ADD,RS}}. In the case of `flat' extra dimensions, \eg, in the model of 
Arkani-Hamed, Dimopoulos and Dvali (ADD){\cite {ADD}}, the 4-d Planck and fundamental scales are related by the volume of the 
compactified extra dimensions: 
\begin{equation}
\mpl^2=V_nM_*^{n+2}\,,
\end{equation}
where $V_n$ is the volume of the compactified manifold. Assuming for simplicity that these extra dimensions form an 
$n$-dimensional torus, if all compactification radii ($R_c$) are the same, then $V_n=(2\pi R_c)^n$. In such a scenario 
gravity becomes strong at $M_*$ and not at $\mpl$ which is viewed as an artifact of our inability to probe gravity 
at scales smaller than $R_c$. This scenario has gotten a lot of attention over the last few years and the collider 
phenomenology of these types of models has been shown to be  particularly rich{\cite {GRW}}. In such a scheme it would 
be natural that the NC scale, $\Lambda_{NC}$, would now also be of order $M_* \sim$ TeV allowing it to be  
accessible to colliders. Furthermore, the copious production of TeV-scale BH at colliders also becomes possible{\cite {BH}} 
and the nature of such BH could then be examined experimentally in some detail. Thus it is reasonable to ask if the properties 
of such TeV-scale BH may be influenced by NC effects, which originate at a similar scale, and if these effects are 
large enough to be observable in collider data. 

The goal of the present paper is to begin to address these issues. In particular we will examine how NC BH in $D$ dimensions 
differ from those in 4-d as well as from the more conventional $D$-dimensional commutative Schwarzschild BH traditionally 
analyzed at colliders. Furthermore, we will demonstrate that the essential features of this scenario are not particularly 
sensitive to the detailed nature of the NC smearing.

\section{Analysis and Results}

We begin our analysis by reminding the reader that we will assume that $D=4+n$-dimensional gravity, and BH in particular, 
can still be described by the conventional Einstein-Hilbert (EH) action, \ie, 
\begin{equation}
S={M_*^{n+2}\over {2}}\int d^{4+n}x ~\sqrt {-g}~R\,,
\end{equation}
with $R$ being the Ricci scalar and $M_*$ being the (reduced) fundamental scale as appearing in the ADD relationship 
above. It is important to recall that for the ADD scenario with $n\geq 2$, $M_*$ is only weakly constrained by 
current collider experiments, \ie, one finds that $M_* \geq 0.38-0.60$ TeV, depending on the value of $n$, when the 
bound on the more commonly used GRW{\cite {GRW}} parameter $M_D>1.5$ TeV is employed. 

In the present and NSS approaches the basic effect of NC is proposed to be the smearing out of 
conventional mass distributions. Thus, following 
NSS{\cite {Nicolini:2005vd}}, we will take, instead of the point mass, $M$, described by a $\delta$-function 
distribution, a static, spherically symmetric, Gaussian-smeared matter source whose NC `size' is determined 
by the parameter $\sqrt {\theta}\sim \Lambda_{NC}^{-1}$:
\begin{equation}
\rho={M\over {(4\pi \theta)^{(n+3)/2}}} e^{-r^2/4\theta}\,. 
\end{equation}
Here will we explicitly assume that both the horizon size of our BH and the NC parameter $\sqrt \theta$ are 
far smaller than the compactification scale $R_c$ so that the BH physics is not sensitive to the finite 
size of the compactified dimensions. For ADD-type models this can be easily verified from the numerical results 
we obtain below as the BH horizon size will typically be of order $\sim 1/M_*$ while $R_c$ is generally many 
orders of magnitude larger as long as the value of $D$ is not too large{\cite{Hewett:2005iw}} as is certainly the case 
when $D \leq 11$.  
Note that the value of $\sqrt \theta$ is directly correlated with the NC scale and is certainly proportional to it; 
however, within this treatment the exact nature of this relationship is unspecified and would require a more detailed 
model to explicitly determine. It is sufficient to remember only that $\theta \approx 1/\Lambda_{NC}$. 
It is important to realize that many such parameterizations of this peaked smeared mass distribution 
are possible which should lead to qualitatively similar physics results. However, the various predictions arising from these 
may differ only at the O(1) level or less, as long as the detailed structure of the peaked mass distribution is not probed. 
We will discuss this issue further below.  

The metric of our $D$-dimensional space is assumed to be given by the usual $D$-dimensional Schwarzschild form 
\begin{equation}
ds^2=e^\nu dx_0^2-e^\mu dr^2-r^2 d\Omega_{D-2}^2\,.
\end{equation}
Here we will be searching for Schwarzschild-like, 
spherically symmetric and static solutions with $\nu$ and $\mu$ being functions only 
of the co-ordinate $r$ and we will further demand that $e^{\nu,\mu} \to 1$ as $r \to \infty$; this 
will require that $\nu=-\mu$ in the solutions of Einstein's equations as in the usual commutative scenario since the EH 
action and resulting field equations remains applicable. Note that the surface 
of the sphere, $\Omega_{D-2}$, can be simply described by a set of $D-2=n+2$ angles, $\phi_i$, where $i=1,..,n+2$.  

Given the assumed form of the matter density, $\rho$, above and the expectation that $\nu=-\mu$ as usual, 
two components of the 
diagonal stress-energy tensor, $T^{AB}$, are already determined, \ie, $T_r^r=T_0^0=\rho$. As noted by NSS, the 
remaining components, $T_i^i$ (no summation), 
all $n+2$ of which are identical due to the spherical symmetry, can be obtained from the requirement that 
$T^{AB}$ have a 
vanishing divergence, $T^{AB};_B=0$, where the semicolon denotes covariant differentiation.  Given the nature of 
our metric it is easily seen that both $T^{00};_0=0$ and $T^{ii};_i=0$, for all $i$, automatically. The remaining 
equation $T^{rr};_r=0$ then yields the explicit result 
\begin{equation}
0=\partial_r T_r^r+{1\over {2}} g^{00}(T_r^r-T_0^0)~\partial_r g_{00}+{1\over {2}}\sum_i g^{ii}(T_r^r-T_i^i)
~\partial_r g_{ii}\,.
\end{equation}
Since by construction $T_r^r=T_0^0$ and, noting that $g^{ii}\partial_r g_{ii}=2/r$, for all (unsummed) $i$, this yields 
\begin{equation}
T_i^i=\rho +{r\over {n+2}}\partial_r \rho\,,
\end{equation}
for all $i$ (without summation). This reproduces the 4-d NSS result in the limit when $n\to 0$.

With our metric the non-zero components of the Ricci tensor are given by (with the index $i$ not summed) 
\begin{eqnarray}
R_0^0=R_r^r &=& -{e^\nu \over {2}} \Bigg[\nu''+(\nu')^2+(n+2){\nu'\over {r}}\Bigg]\nonumber \\
R_i^i &=&{-1\over {r^2}}\Bigg[e^\nu(1+n+r\nu')-(n+1)\Bigg]\,,
\end{eqnarray}
where now a prime denotes partial differentiation with respect to $r$. The Einstein equations resulting from the EH action 
augmented with our matter distribution can be conveniently written in the form  
\begin{equation}
R_A^B={1\over {M_*^{n+2}}}\Bigg(T_A^B-\delta_A^B {T\over {n+2}}\Bigg)\,,
\end{equation}
where $T$ is the trace of the stress-energy tensor, $T=T_A^A$. Note that given our assumptions there are only two 
distinct Einstein equations. Writing $g_{00}=e^\nu=1-A(r)$, the $R_i^i$ Einstein 
equation leads to the following first order differential equation for $A(r)$: 
\begin{equation}
A'+{{n+1}\over {r}}A={1\over {M_*^{n+2}}}{{2r\rho}\over {n+2}}\,,
\end{equation}
from which we obtain the following solution, after 
substituting the above expression for $\rho$ and demanding that $A(r)\to 0$ as $r\to \infty$:  
\begin{equation}
A(r)={1\over {M_*^{n+2}}} {M\over {(n+2)\pi^{(n+3)/2}}} {1\over {r^{n+1}}}\int_0^{r^2/4\theta} dt~e^{-t}t^{(n+1)/2}\,. 
\end{equation}
This result is seen to reduce to that of NSS when $n\to 0$ as well as to the usual $D$-dimensional Schwarzschild 
solution when $\theta \to 0$. The remaining $R_0^0$ Einstein equation just returns to us 
the continuity equation for the stress-energy tensor, hence, nothing new. Note that given our assumptions these results 
allow us to calculate $A(r)$ for any chosen form of the mass distribution $\rho(r)$.

The horizon radius, $R_H$, occurs at values of $r$ where $g_{00}=0$, \ie, where $A(r)=1$. Defining for convenience 
the dimensionless quantities $m=M/M_*$, $x=M_*R_H$ and $y=M_*\sqrt{\theta}$, along with the constant 
\begin{equation}
c_n={{(n+2) \pi^{(n+3)/2}}\over {\Gamma({{n+3}\over {2}})}}\,,
\end{equation}
the horizon radius can be obtained by solving the equation 
\begin{equation}
x^{n+1}={m\over {c_n}}F_n(z)\,,
\end{equation}
where $z=x/y$ and the functions $F_n(z)$ are given by the integrals
\begin{equation}
F_n(z)={1\over \Gamma({{n+3}\over {2}})}\int_0^{z^2/4} dt~e^{-t}t^{(n+1)/2}\,.
\end{equation}
These integrals be performed analytically when $n$ is odd, \eg, $F_1=1-e^{-q}(1+q)$, $F_3=1-e^{-q}(1+q+q^2/2)$ 
and $F_5=1-e^{-q}(1+q+q^2/2+q^3/6)$, \etc, with $q=z^2/4$. (For $n$ even these functions can be expressed in 
terms of combinations of error functions.) Given their definition it is clear that the $F_n$ 
vanish as $\sim z^{n+3}$ when $z\to 0$ and they are seen to monotonically increase with increasing $z$; as $z \to \infty$ 
our normalization is such that $F_n \to 1$. This implies 
that the BH mass, $m$, diverges as either $x\to 0,\infty$ for fixed $y$ 
and that a minimum value of $m$ must exist for some value of $x$. 
Since $x$ appears on both sides of the above equation 
determining the horizon radius, a trivial relationship between $m$ and $x$ no longer 
occurs as it does for the ordinary $D$-dimensional Schwarzschild BH solution. 
In the $\theta,y \to 0$ limit, corresponding to the usual commutative result, the upper limit of the integral 
defining the functions $F_n$ becomes infinite and we arrive at the well-known 
standard result{\cite {BH}} as then $F_n \to 1$:  
\begin{equation}
m=c_nx^{n+1}\,.
\end{equation}

Note that if we had chosen a different form for the matter distribution representing the smeared point mass source 
only the set of functions $F_n(z)$ would be changed but their general properties would be identical to those above. 
For example, if we had taken a smearing of the modified Lorentzian form, $\rho \sim (r^2+\theta'^2)^{-(n+4)/2}$, 
(with the parameter 
$\theta'$ not necessarily being the same as $\theta$) then the corresponding functions, which we'll call $G_n$, would 
also vanish as $z\to 0$ in a power law manner and go $\to 1$ as $z\to \infty$ in a monotonic fashion. For example, one obtains 
$G_0(z)={2\over {\pi}}(\tan^{-1}z-{z\over {1+z^2}})$ which we observe has these same limiting properties and is quite 
similar to the $F_n$ qualitatively. More generally we find that the $G_n$ are given by the integrals 
\begin{equation}
G_n(z)={2\over {\pi}}{{(n+2)!!}\over {(n+1)!!}}\int_0^z ~dt ~{{t^{n+2}}\over {(1+t^2)^{(n/2+2)}}}\,.
\end{equation}
These basic properties of the $F_n$ (and $G_n$) capture the essential aspects of the NC physics. 
Other possible smearings, such as a straightforward exponential, would also 
lead to quite similar results.{\footnote {For the case of an 
exponential, the corresponding functions, $Q_n(z)$, are found to be related to the $F_n$ above as 
$Q_n(z)=F_{2n+3}(z^2/4)$.}}  We will have more to say about this below but note that many of the specific expressions 
that we will obtain remain applicable if we take a different form for the smeared mass distribution. 

It is interesting to inquire how the NC value of $x$, for the moment explicitly denoted as $x_{NC}$, compares with 
the usual $D$-dimensional result. Clearly for any fixed value of $m$ the ratio $x_{NC}/x$ 
can be expressed solely via the functions $F_n$ and is thus only dependent upon the ratio $z=x/y$ and $n$; the result of 
this calculation is shown in Fig.~\ref{fig1}. In examining these results, as well as those in the following figures, it is 
important to remember than when $n=0$, $M_*=\mpl$; for all larger values of $n$, $M_* \sim 1$ TeV.  
In this figure we see that for large values of $z\gsim 3$ we recover the commutative result for all $n$ 
since the NC scale is far smaller than the horizon size in this case. However when the two scales are comparable or 
when the horizon shrinks inside the NC scale the value of $x$ is greatly reduced for fixed $m$. Of course, we really 
should not 
trust the details of our modeling of the NC effects when $z$ is extremely small, \ie, when $x$ is very much less than $y$. 
It is also in this very region where most of the differences between, \eg, the Gaussian and Lorentzian forms of the 
smeared mass distribution would be expected to begin to appear. 
\begin{figure}[htbp]
\centerline{
\includegraphics[width=8.5cm,angle=90]{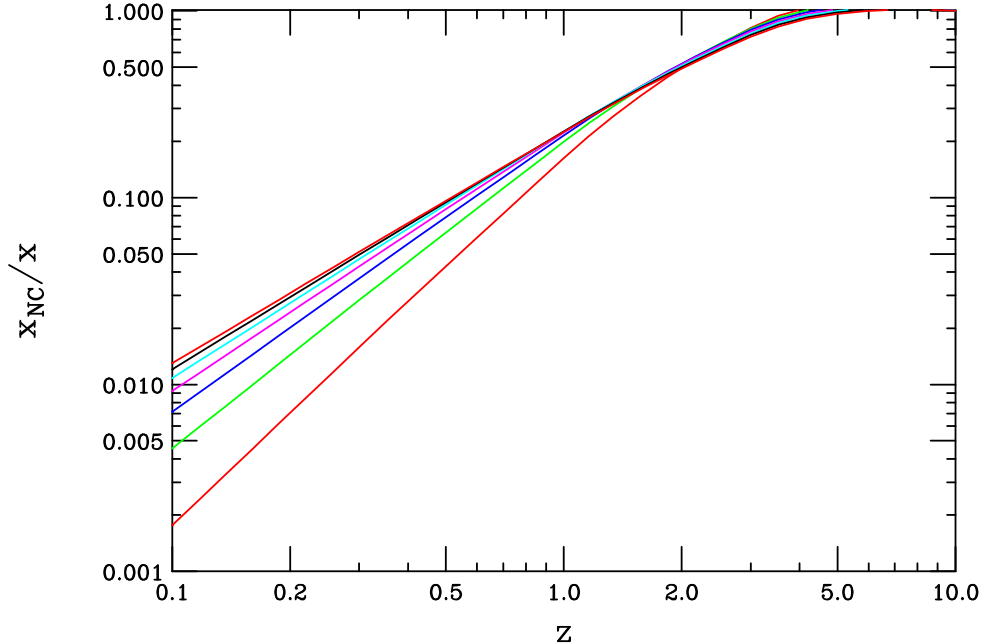}}
\vspace*{0.1cm}
\caption{Horizon size in the NC scenario compared to the commutative result as a function of $z=x/y$ for fixed values 
of $m$. On the left-hand side of the figure, from bottom to top, the curves correspond to $n=1$ to 7.}
\label{fig1}
\end{figure}

\begin{figure}[htbp]
\centerline{
\includegraphics[width=8.5cm,angle=90]{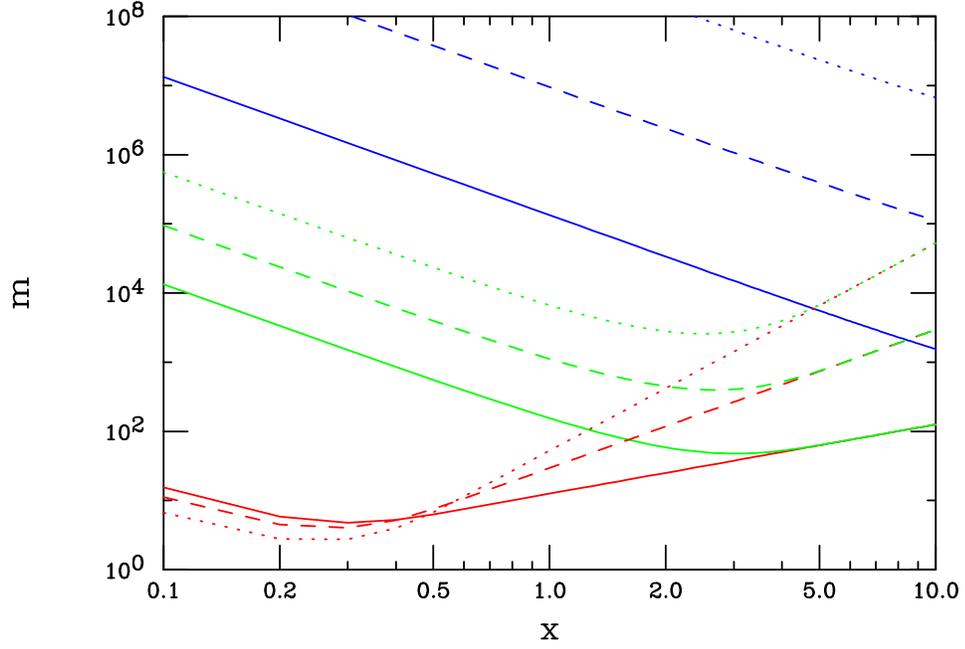}}
\vspace*{0.1cm}
\caption{BH mass as a function of the horizon size for $n=0$(solid), $n=1$(dashed) and $n=2$(dotted). The 
upper(middle, lower) set of curves correspond to $y=10(1,0.1)$.}
\label{fig2}
\end{figure}

More generally, we can now calculate $m$ as a function of $x$ for fixed $y$. Instead of a monotonically increasing 
function $\sim x^{n+1}$, we find that $m(x)$ is now a function with a single minimum and which grows large as either 
$x\to 0$ or $\to \infty$. For small $x$ we find the scaling $m\sim y^{n+3}/x^2$ while the usual commutative behavior  
is obtained at large $x$, $\sim x^{n+1}$. 
Fig.~\ref{fig2} gives a first rough indication of this behavior. The existence of a minima  
has several implications: ($i$) a minimum value of $m$ implies that there is a physical mass threshold below which 
BH will not form. ($ii$) The inverse function, $x(m)$, is double valued indicating the existence of two possible  
horizons for a fixed value of the BH mass; this is a potentiality first pointed out by NSS and something we will return to 
below. In Fig.~\ref{fig2} we see several additional  
features: first, as $y$ increases for fixed $n$ so does $m$ except where $x$ is large and we are thus residing in the 
commutative limit. Also for large $y$ we see that $m$ increases with $n$ as it usually does in the 
commuting case. This is not too surprising given the small $x$ scaling behavior of $m$ above. 
Secondly, we see that for large $y$ the values $m$ are always large and the position of the mass minimum moves out 
to ever larger values of $x$ as $y$ increases. For example, if $y=n=1$ then $m\gsim 400$ and lighter 
BH do not form. Such large mass values are far beyond the range accessible to the LHC if we assume $M_* \sim 1$ TeV, 
\ie, the interesting range roughly being $1 \lsim m \lsim 10$ or so.{\footnote 
{Note that we do not expect BH to form for masses much less than $M_*$ so that it is reasonable to believe 
that $m \gsim 1$.}} 

\begin{figure}[htbp]
\centerline{
\includegraphics[width=8.5cm,angle=90]{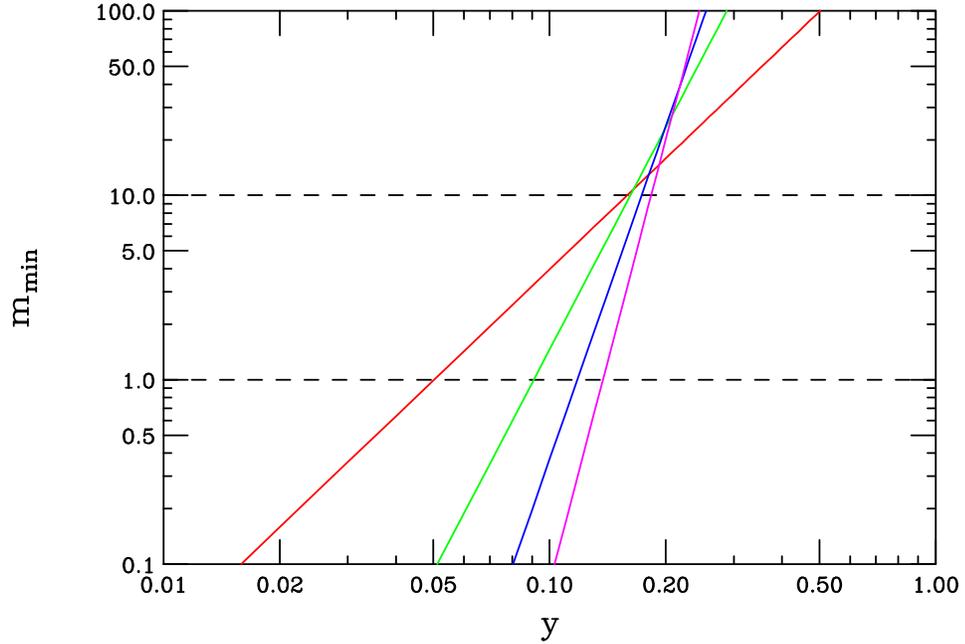}}
\vspace*{0.1cm}
\caption{Minimum BH mass as a function of $y$ showing the (very) approximate mass range accessible to the LHC between the 
dashed lines. On the left-hand side from top to bottom the curves correspond to $n=1,3,5$ and 7, respectively. The 
allowed parameter range is above and to left of each curve.}
\label{fig0}
\end{figure}

Since we are 
interested here in BH that can be created at colliders we will restrict our attention to smaller $y$ values. In fact 
a short calculation shows that we need $0.05\lsim y\lsim 0.2$ in order to get into the LHC accessible mass region 
$1 \lsim m \lsim 10$. To demonstrate this we must find the minimum value of the BH mass, $m_{min}$ as a function of $y$ 
for various values of $n$. This can be done in a two-step procedure: first we find where in $x$ the mass minimum 
occurs for fixed values of $y$, \ie, where $\partial m/\partial x=0$. Calculating this derivative we see that  
the minimum can be obtained by solving the equation (using for convenience the variable $q=x^2/4y^2$ introduced above):  
\begin{equation}
F_n(q)-{{2q^{(n+3)/2}e^{-q}}\over {(n+1)\Gamma({{n+3}\over {2}})}}=0\,,
\end{equation}
which has a single, non-zero root $q_0(n)$. For any $y$ this tells us the value of the horizon radius where the minimum 
mass occurs, $x_{min}=2y\sqrt {q_0(n)}$, which we can now use to obtain $m_{min}$ employing 
the equations above. The result of 
this calculation is shown in Fig.~\ref{fig0}. Here we see that for $n$ in the range 1 to 7 the relevant values of $y$ 
are rather narrow, not differing from $y\simeq $0.1 by more than about a factor of 2.

\begin{figure}[htbp]
\centerline{
\includegraphics[width=8.5cm,angle=90]{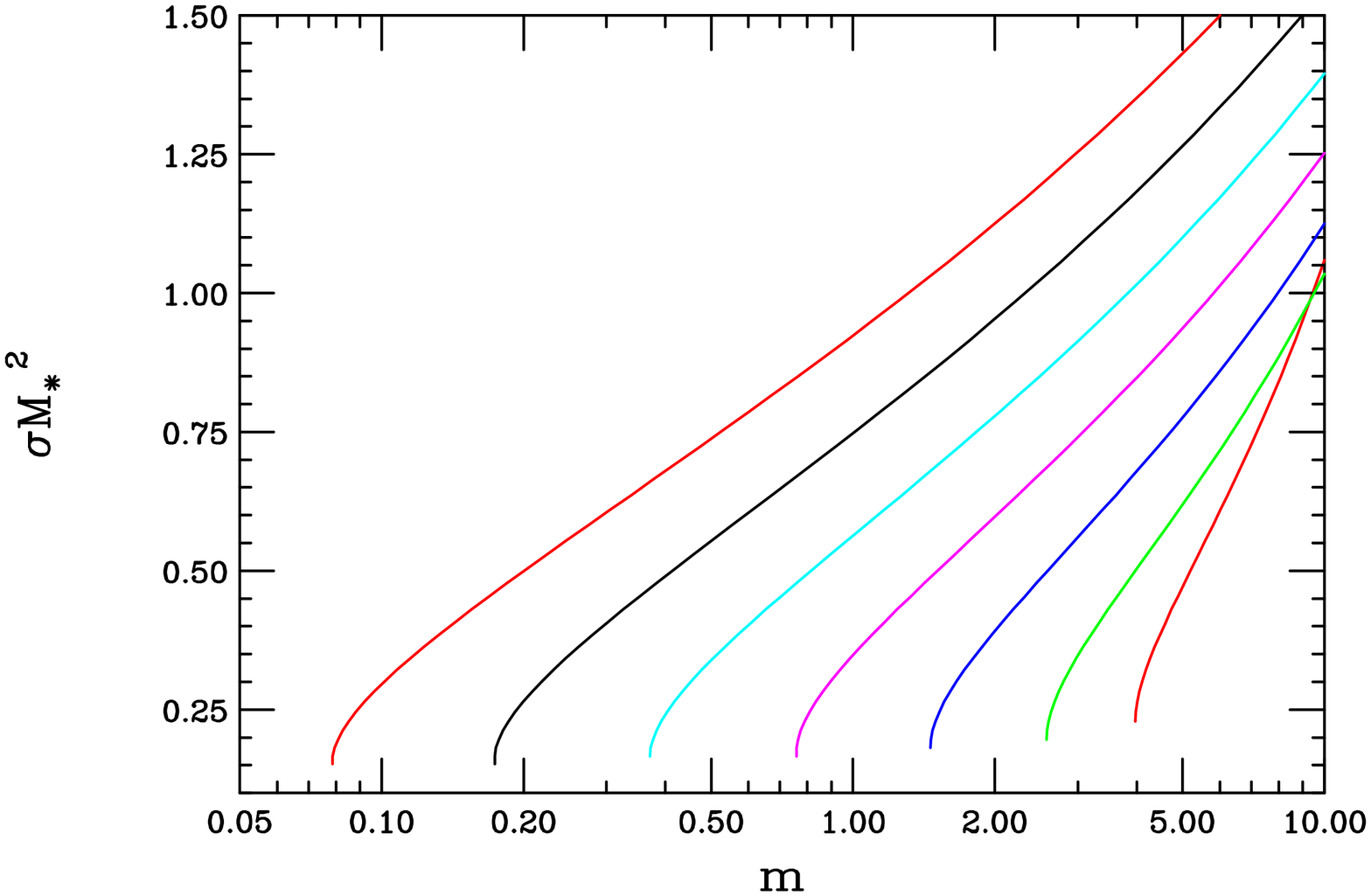}}
\vspace*{0.1cm}
\caption{NC BH production subprocess cross section as a function of $m$ for $y=0.1$. From right to left at the bottom of the 
figure the curves correspond to $n$=1 to 7.}
\label{figy}
\end{figure}

It is important to recall that for the ordinary commutative $D$-dimensional BH solution both 
$m_{min}$ and $x_{min}$ are algebraically zero. However, since we {\it believe} that $m_{min}\gsim 1$ is required to produce 
a BH, stronger conditions are usually imposed. In our case 
the results shown in Fig.~\ref{fig0} represent the {\it algebraic} lower bound on $m$ which certainly $\to 0$ as 
$\theta \to 0$. Physically, we might crudely imagine that $m_{min}\simeq Max(1,m_{min}^a)$ where $m_{min}^a$ is the result 
shown in Fig.~\ref{fig0}. We note that having a finite $m_{min}$ implies a BH production threshold while a finite 
$x_{min}$ implies a minimum BH production cross section, $\sigma_{BH} \simeq \pi x_{min}^2/M_*^2$, at colliders such as 
the LHC as is shown in Fig.~\ref{figy}. Here we see that far above threshold the BH production cross section scales like  
$\sim m^{2/(n+1)}$ as would be expected in the commutative theory. However, for lighter BH this cross section falls 
significantly below this simple scaling rule and becomes quite small in the neighborhood of $m_{min}$, almost but not quite 
vanishing.  


\begin{figure}[htbp]
\centerline{
\includegraphics[width=7.5cm,angle=90]{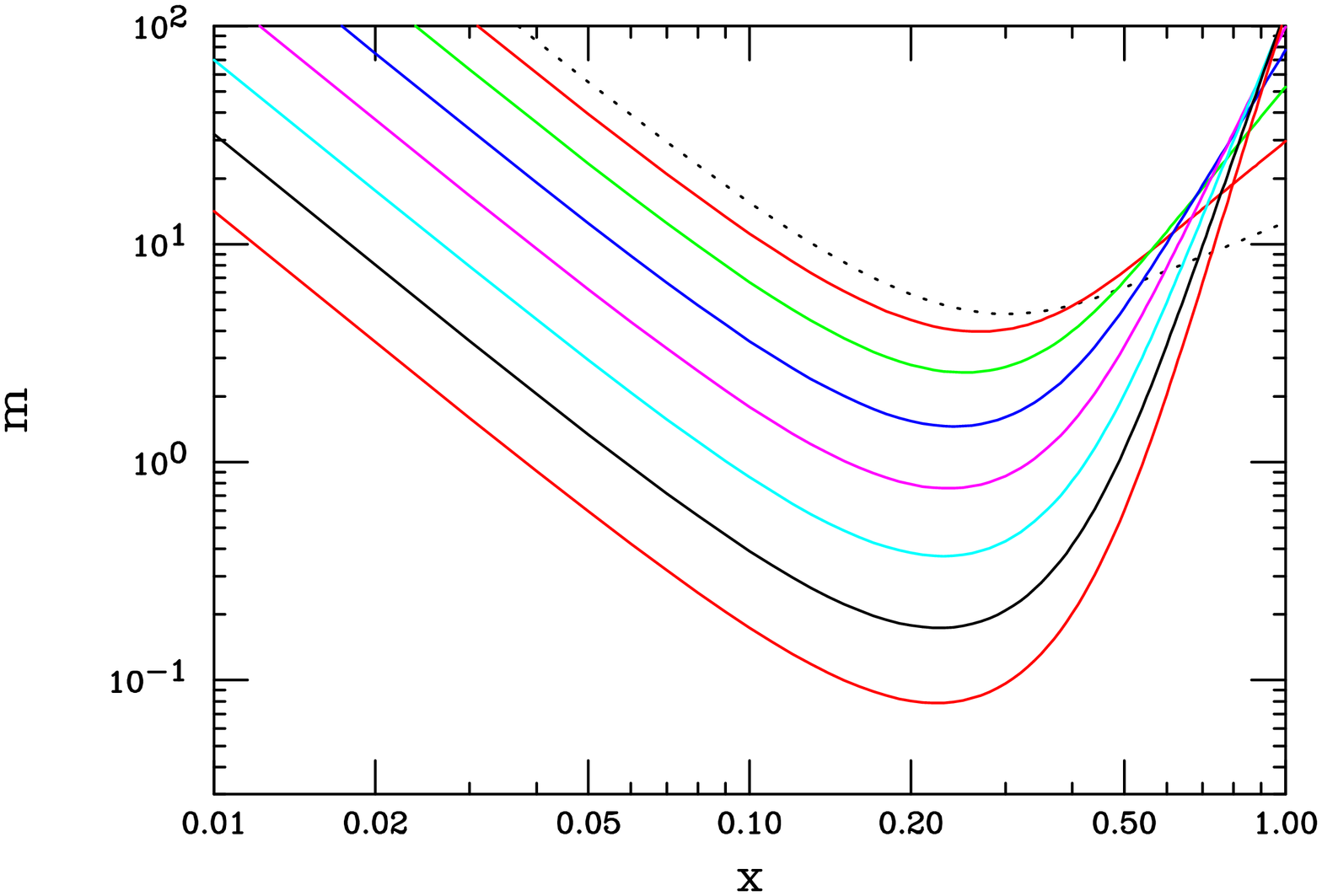}}
\centerline{
\includegraphics[width=7.5cm,angle=90]{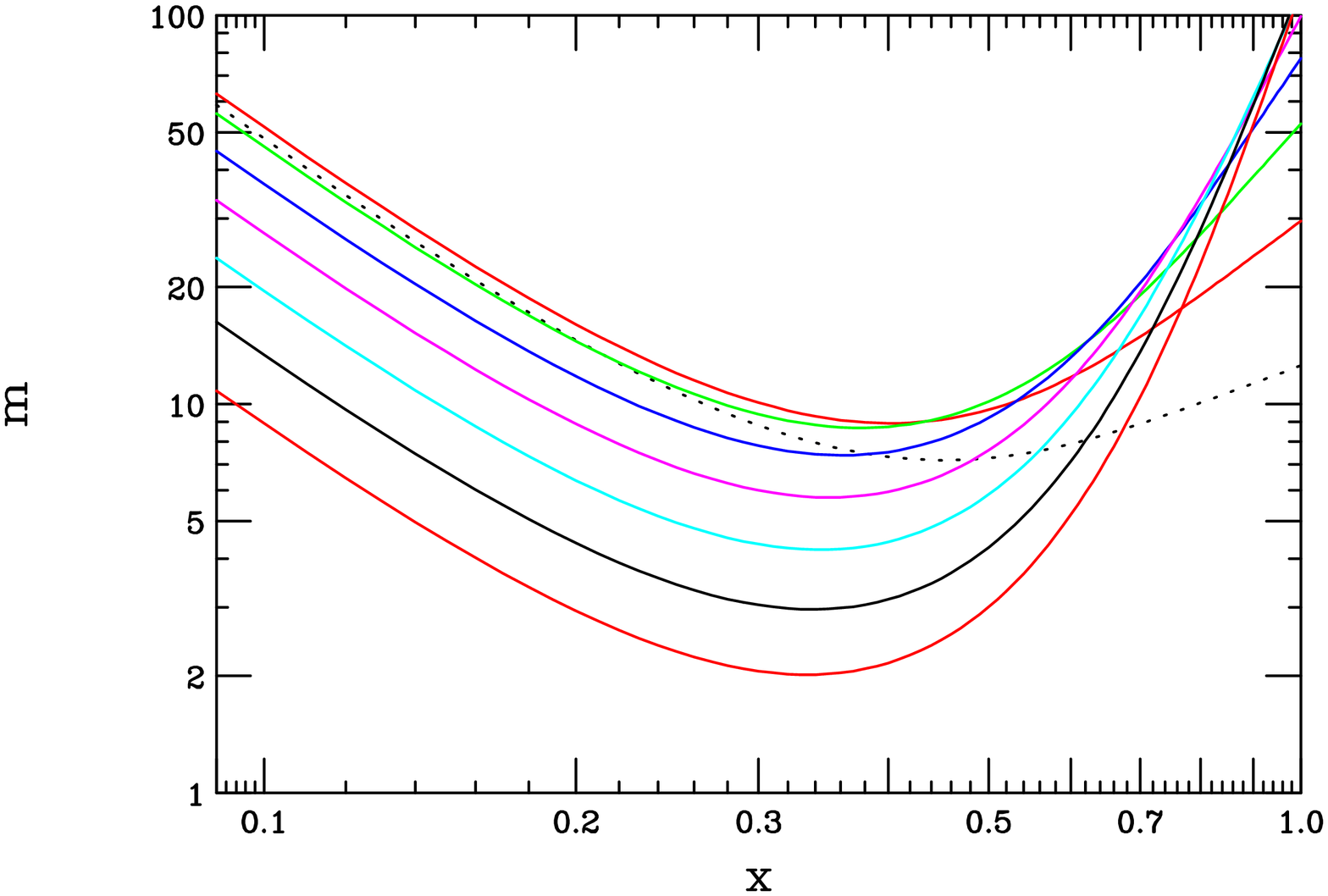}}
\centerline{
\includegraphics[width=7.5cm,angle=90]{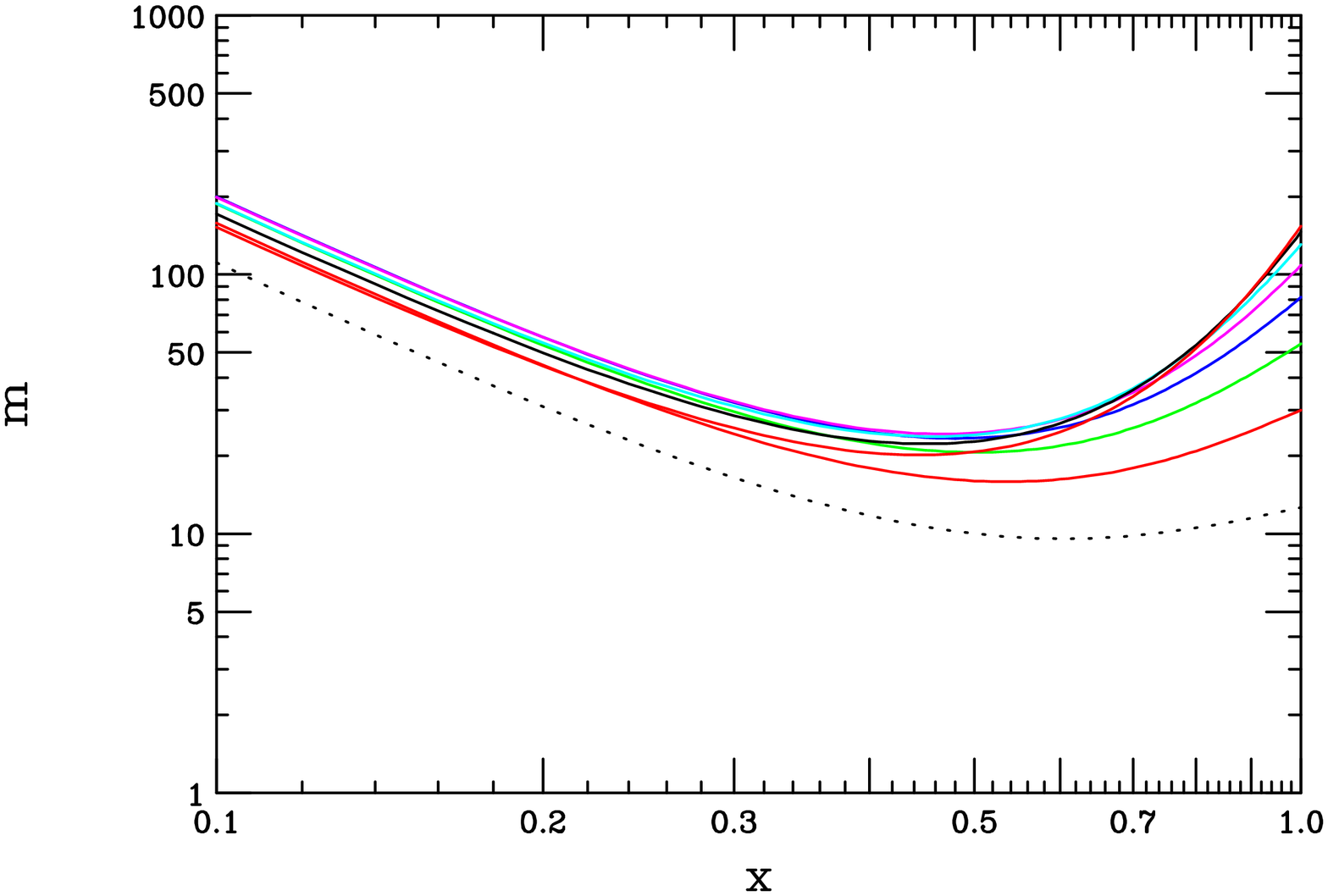}}
\vspace*{0.1cm}
\caption{Same as in Fig.~\ref{fig2} but now for $y=0.10$(top), 0.15(middle) and 0.20(bottom) with the dotted 
curve being for $n=0$. From top to bottom on the left-hand side of the figure the solid curves are for $n$=1 to 7.}
\label{fig3}
\end{figure}

We can now focus on this region of small $y$; for simplicity in what follows we will generally 
concentrate our results on the case $y=0.1$.   Fig.~\ref{fig3} shows $m(x)$ for $y=0.1-0.2$ 
for $n$ in the range $0$ to 7. 
Note that {\it generally} larger $n$ leads to smaller horizon radii for fixed $y$ and already at $y=0.2$ we see 
explicitly that BH are too massive to be produced at the LHC as expected from the above analysis if $M_* \sim 1$ TeV. 
One also sees that at $x \sim 1$ the asymptotic behavior,  $m \sim x^{n+1}$, has already begun to set in for all $n$. 

\begin{figure}[htbp]
\centerline{
\includegraphics[width=8.5cm,angle=90]{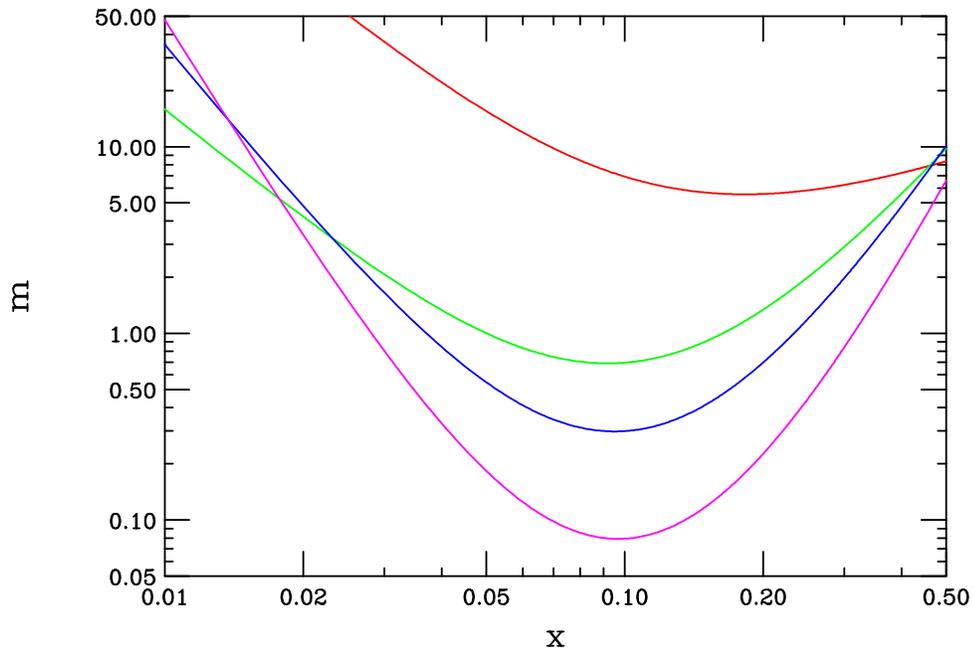}}
\vspace*{0.1cm}
\caption{BH mass as a function of the horizon size $x$ for $y'=0.1$ using Lorentzian smearing. From top to bottom in 
the middle of the figure the curves correspond to $n=0$, 2, 4 and 6, respectively.}
\label{fig3p}
\end{figure}

As an example of the insensitivity of these results to our  
Gaussian parameterization of the smearing due to NC effects, let us 
briefly considered the modified Lorentzian form mentioned above. Setting $y'=M_*\sqrt {\theta'}=0.1$ (which is not 
necessarily the same as $y=0.1$), we again evaluate 
$m$ as a function of $x$. The result of this calculation is shown in Fig.~\ref{fig3p} which we should compare with 
the top panel of Fig.~\ref{fig3} that yields essentially the same result in the asymptotic large $x \gsim 0.5$(as it 
should since this is 
the commutative limit). In both cases a minimum mass occurs at relatively small $x$ with somewhat similar values of $m$. 
$m_{min}$ decreases in both cases as $n$ is increased and the corresponding value of $x_{min}$ also decreases as $n$ is 
increased. The $m_{min}$ values are seen to be quite comparable in the two cases. The greatest difference in the two 
results is seen to occur in the region of $x$ below $x_{min}$ where there is the most sensitivity to NC effects and 
the detailed shape of the BH mass distribution. This is just what we would have expected; in the region where 
NC effects just begin to be felt the detailed nature of the peaked mass distribution is not actually 
being probed. What is really 
being probed in this parameter range is the fact that there is a peaked mass distribution instead of a $\delta$-function 
source, \ie, the BH has a form factor due its finite size, and not the details of its shape. Only at smaller values of the 
radii (relative to the values of $y$ or $y'$) do the differences in the details of the mass distribution become important.  
In fact, we can if we wish tune our chosen value of $y'$ to make these two sets of curves even more alike. 
Since the majority of the effects we will 
discuss are sensitive only to physics with $x \ge x_{min}$, this short analysis shows that the general 
features of the results that we obtain below are not particularly sensitive to how the NC smearing of the mass 
distribution is performed. 

\begin{figure}[htbp]
\centerline{
\includegraphics[width=8.5cm,angle=90]{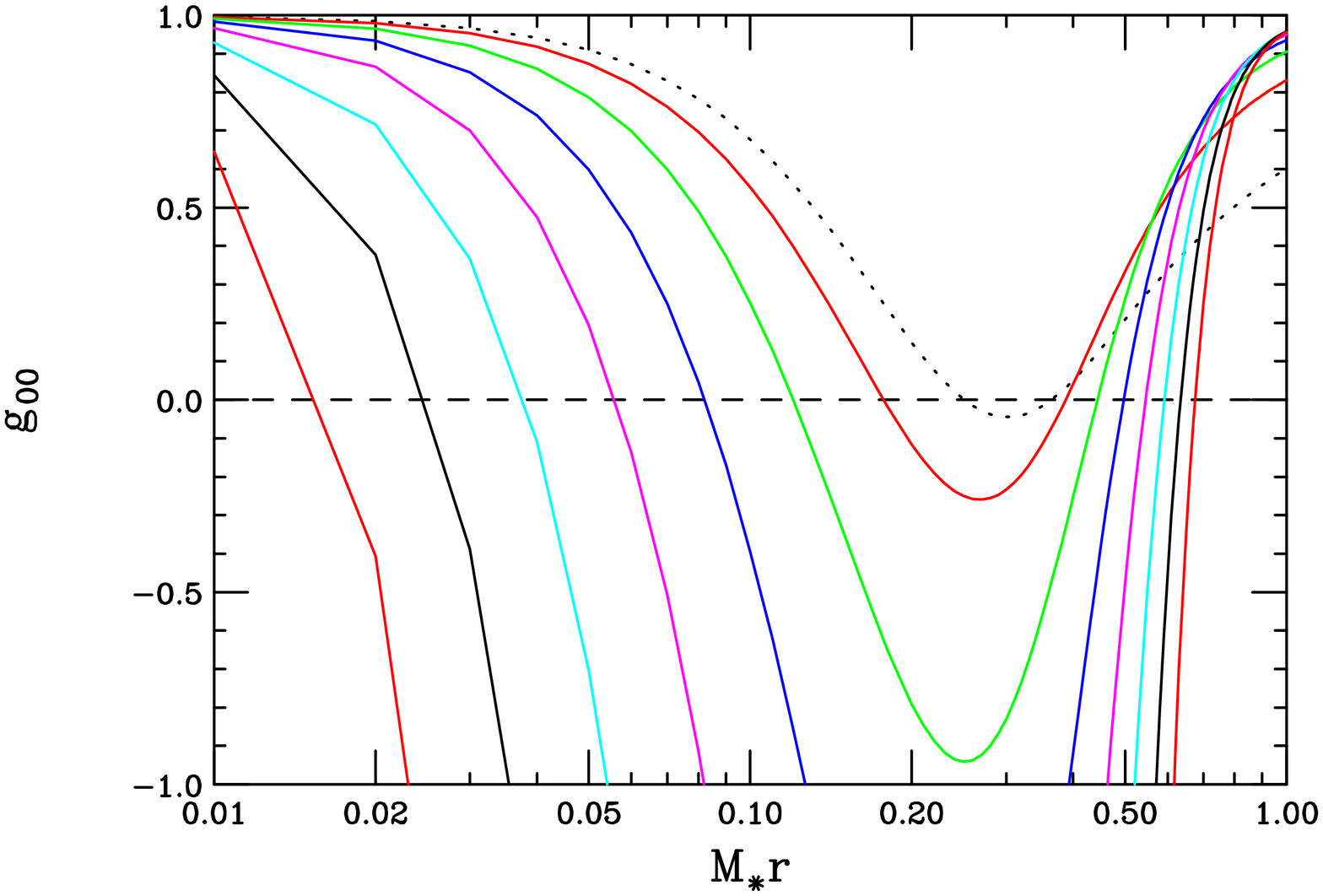}}
\centerline{
\includegraphics[width=8.5cm,angle=90]{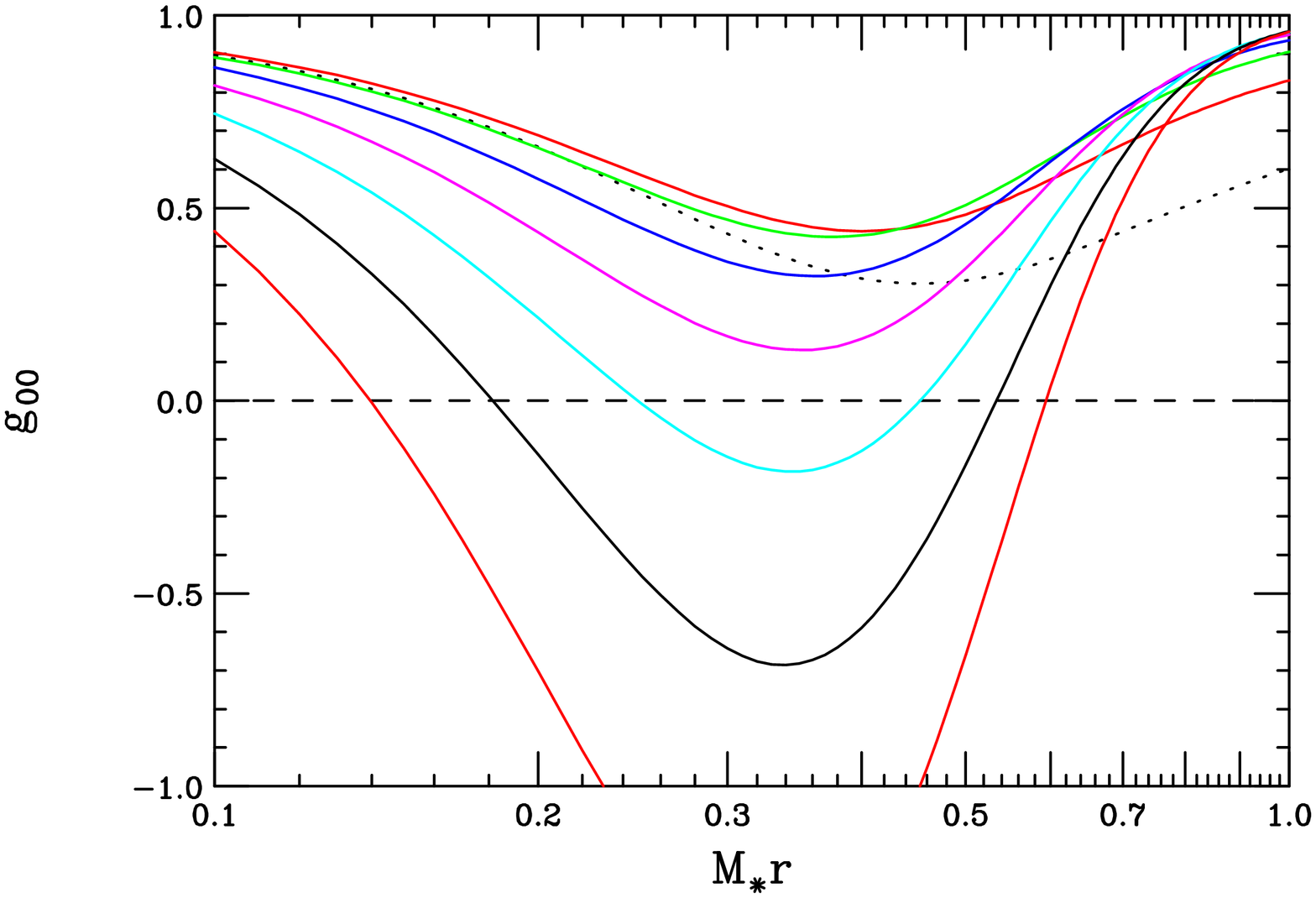}}
\vspace*{0.3cm}
\caption{$g_{00}$ as a function of $M_*r$ assuming $m=5$ for either $y=0.10$(top) or 0.15(bottom). The dotted curve 
corresponds to $n=0$ while from top to bottom on the left-hand side of the figure the solid curves are for $n$=1 to 7. 
The dashed line corresponds to $g_{00}=0$.}
\label{fig4}
\end{figure}

In order to understand the possibility of the formation of two (or no) horizons, we follow NSS and consider the 
metric tensor component 
$g_{00}$ as a function of the dimensionless radius co-ordinate $M_*r$ as shown in Fig.~\ref{fig4}. Here we will assume 
that $m=5$, a typical value which is kinematically accessible at LHC, for demonstration purposes. Recall that 
horizons occur when $g_{00}=0$. With $y=0.1$ all of the curves pass through $g_{00}=0$ twice 
corresponding to two horizons, one on either side of $x_{min}$. This explains why $x(m)$ is double 
valued in  Fig.~\ref{fig3}, \ie, 
the two solutions correspond to the two radii where $g_{00}$ vanishes. For $n=0$, the case 
studied by NSS, these two horizons are rather close in radius but this separation grows significantly as $n$ increases. 
When $y=0.15$, we see that for $n=0-4$ no horizon form as $ g_{00}\gsim 0.13$. This corresponds to the result observed 
in Fig.~\ref{fig3} for $y=0.15$ where we see that for this range of $n$ the value $m=5$ is not allowed. 
However, for $n \geq 5$, we again 
obtain two horizons; clearly a tuning of parameters will allow the two horizons to converge to the case of 
a single degenerate 
horizon at $x_{min}$ as found by NSS. For both values of $y$ we note that as $M_*r \to 0$ the metric is no longer singular 
as in the pure Schwarzschild case as was noted by NSS in 4-d (as we are inside a well-behaved mass distribution),  
independently of the existence of any horizons. This is further confirmed by 
constructing the Ricci curvature invariant, $R$, as can be easily done from the Einstein equations above; in fact, we 
find that as $x\to 0$, $R \sim y^{-(n+3)}$. Furthermore, and more explicitly, apart from an overall numerical 
factor, $R \sim my^{-(n+3)}[n+4-{{(M_*r)^2}\over {2y^2}}]e^{-(M_*r)^2/4y^2}$ so that $R$ is seen 
to vanish as $M_*r \to \infty$ as expected and 
undergoes a change of sign in the region, \eg, $M_*r \sim 0.2-0.5$ for $y=0.1$ independently of $m$ and only weakly 
dependent on the value of $n$. As we will now see only the `outer' horizon, \ie, the one with $x\geq x_{min}$ is actually 
relevant to us.

\begin{figure}[htbp]
\centerline{
\includegraphics[width=8.5cm,angle=90]{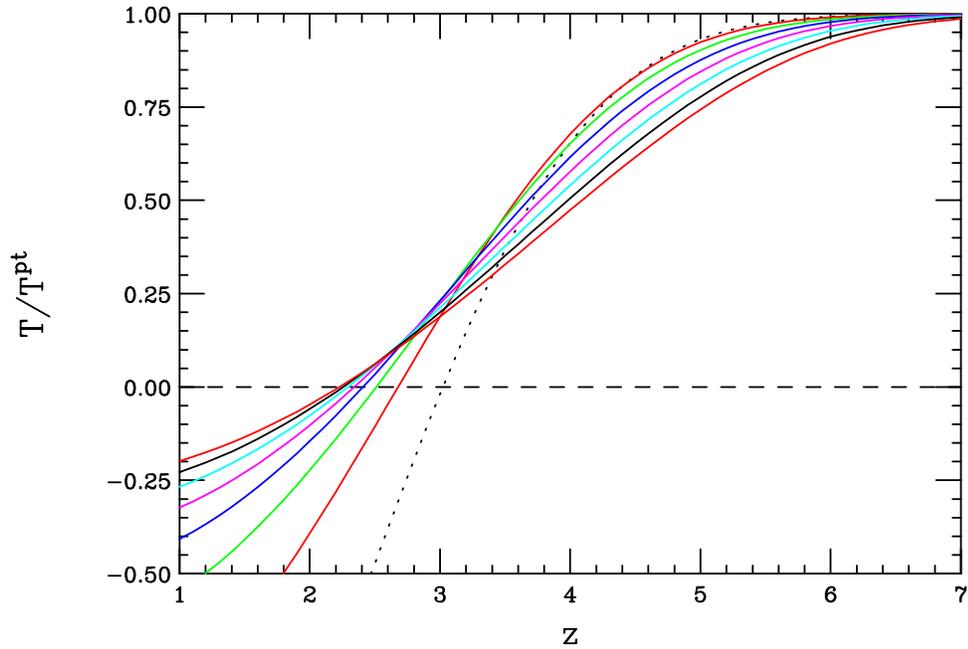}}
\vspace*{0.1cm}
\caption{The ratio of the BH NC temperature to that obtained in the commutative limit where there is only a point matter 
source. The dotted curve corresponds to $n=0$ whereas, from bottom to top on the left-hand side of the figure, the 
solid lines correspond to $n$=1 to 7.}
\label{fig5}
\end{figure}

Our next step is to determine the thermodynamic behavior of these NC BH; to do this we first must calculate the 
Hawking temperature of the BH. This can be done in the usual manner by remembering that 
\begin{equation}
T_H={1\over {4\pi}} {{de^\nu}\over {dr}}|_{r=x/M_*}\,,
\end{equation}
\ie, the temperature is essentially the $r$ (radial) co-ordinate derivative of the metric evaluated at the horizon radius. 
Defining for convenience the dimensionless temperature, $T=T_H/M_*$, we obtain from the above form of the metric   
\begin{equation}
T={{n+1}\over {4\pi x}}\Bigg[1-{{2q^{(n+3)/2}e^{-q}}\over {F_n(q)(n+1)\Gamma({{n+3}\over {2}})}}\Bigg]\,.
\end{equation}
Note that as expected $T$ returns to the usual commutative result in the $q\to \infty$ limit, \ie, the quantity in the 
large square bracket above $\to$ 1 in this limit. It is instructive to 
compare the temperature we obtain in both the NC and commutative cases for various values of the parameters; the ratio 
of these quantities is shown in Fig.~\ref{fig5}. This ratio is seen to be near unity for large $z$ as one would expect 
but it decreases rapidly 
as $z$ approaches the $\sim$2-3 range from above. The temperature is also seen to vanish 
at the same $z$ value where the BH mass is minimized. 
For smaller $z$, $T$ becomes negative (which is where the second horizon occurs) 
and thus we enter a region that might usually be considered unphysical. 
If we had instead chosen the Lorentzian smearing these results would be quite similar qualitatively. 

\begin{figure}[htbp]
\centerline{
\includegraphics[width=8.5cm,angle=90]{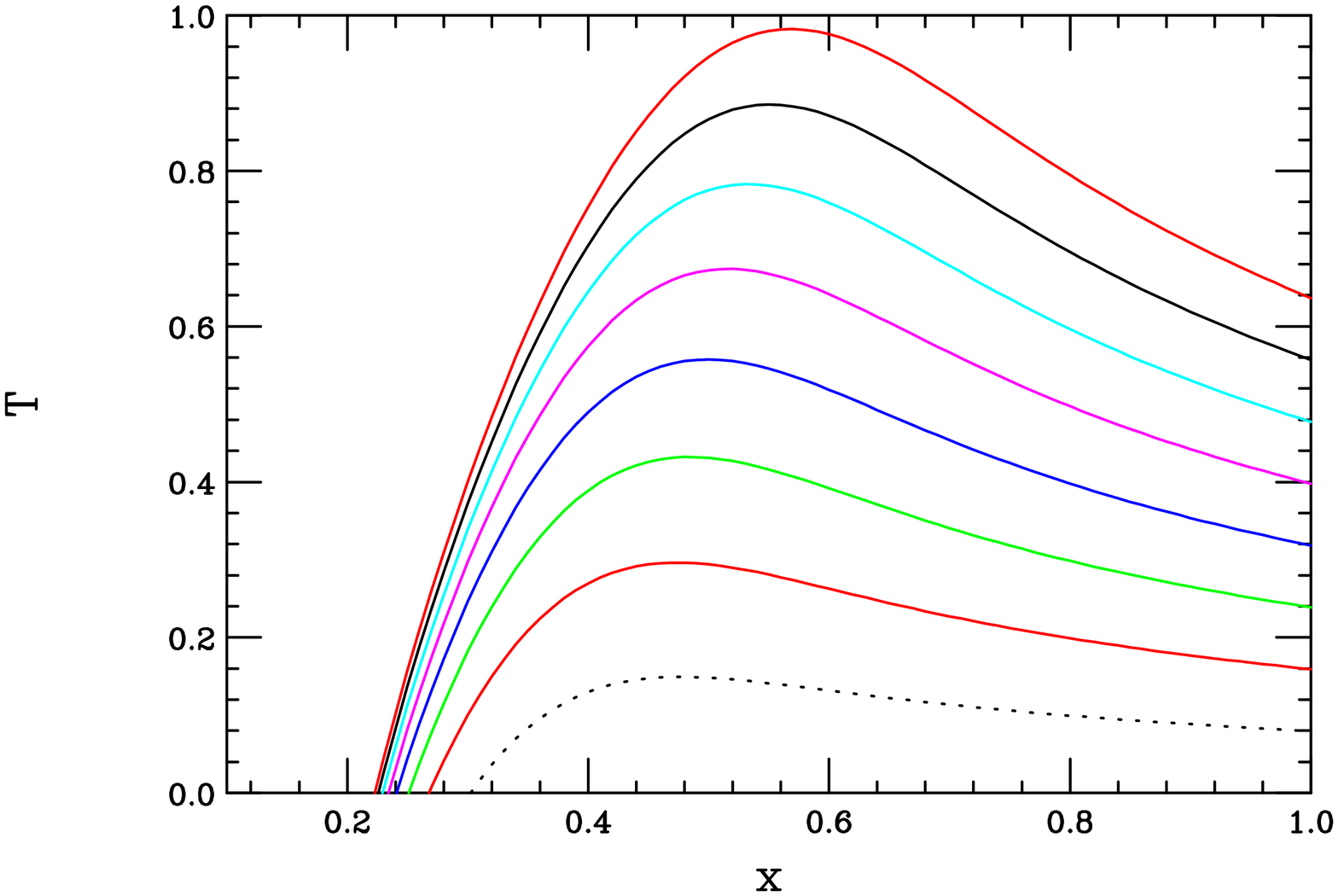}}
\vspace*{0.3cm}
\centerline{
\includegraphics[width=8.5cm,angle=90]{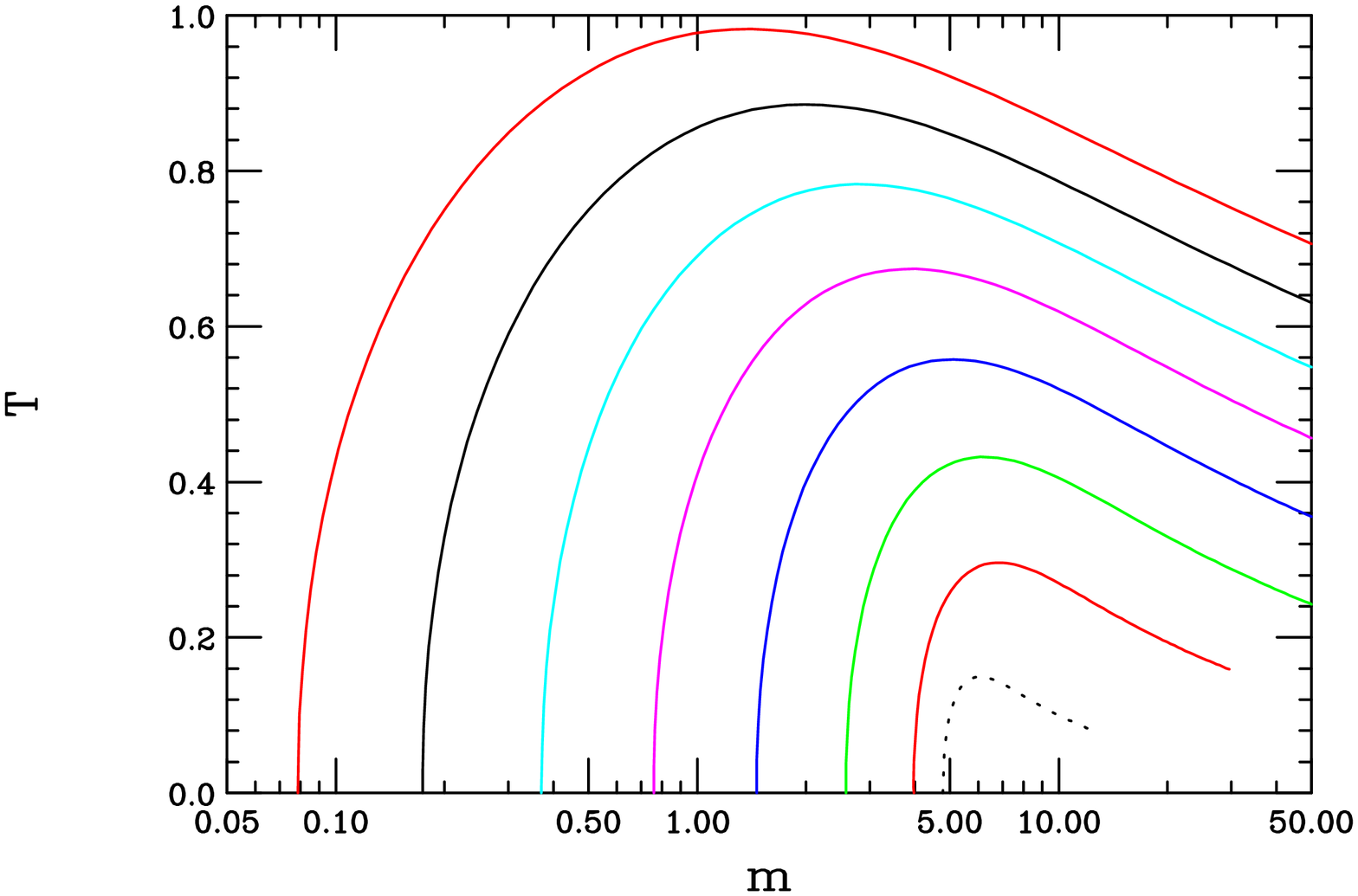}}
\vspace*{0.1cm}
\caption{The NC BH temperature as a function of $x$(top panel) and $m$(lower panel) for fixed $y=0.1$. 
The dotted curve corresponds to case of $n=0$ 
whereas, from bottom to top on the right-hand side of the figure, the solid lines correspond to $n$=1 to 7.}
\label{fig6}
\end{figure}

Fig.~\ref{fig6} shows the actual NC temperature as a function of $x$ for $y=0.1$ and different $n$ values. Here we see 
that $T$ is quite close to its commutative value for $x$ near unity, goes through a maximum as $x$ decreases and then 
falls to zero at the BH mass minimum point. From these results and Fig.~\ref{fig3} above we can now trace the 
history of the entire semiclassical BH evaporation process. 
{\footnote {Before beginning this discussion we must recall the usual argument discussed above that if $m_{min}$ lies 
below $\sim 1$ then no BH will form; we will ignore this prejudice in the following discussion 
treating $m_{min}$ as the true minimum BH formation mass as derived from the theory itself.}}  
Consider a BH formed in a suitable parameter space region with moderately large values of $m\sim 8-10$ 
(and, hence, with large $x$). For such BH their Schwarzschild radii are too large to 
feel the effects of the NC scale in the formation process since $x>>y$. As in the usual 
commuting picture, when the BH emits Hawking radiation it loses mass and gets hotter and thus radiates even more quickly. 
As the BH shrinks it begins to feel the NC effects and the temperature reaches a maximum 
in the mass region  $m\simeq 1-7$, the 
specific number depending on the value of $n$. As the BH continues to lose mass its temperature 
now {\it decreases} so that 
it radiates ever more slowly. Finally, as $m$ approaches $m_{min}$ the semiclassical radiation emission processes ceases 
since $T \to 0$ has been reached leaving a classically stable remnant. 
Though sounding somewhat unusual such a possibility has been discussed in the 
literature for a number of alternative BH scenarios which go beyond the basic picture presented by General 
Relativity based on just the EH action{\cite {BH,remnant}}. {\footnote {These include models with higher curvature 
invariants in the action as well as those where a minimum length scale exists or where the Newton constant is taken to be 
a running parameter.}} Whether quantum effects destabilize such a relic is not known. 

It is easy to convince oneself that a classically 
stable remnant is the natural outcome of this scenario. In the usual treatment of BH 
decays, the mass loss rate (assuming a perfect radiator) is given by 
\begin{equation}
{{dm}\over {dt}}=-\Xi_d x^{d+2}T^{d+4}\,,
\end{equation}
with $\Xi_d$ being a positive 
numerical constant. For decays dominantly to bulk(brane) fields we have $d=n(0)$. In either case, 
clearly the lifetime of the BH is then given by 
\begin{equation}
t_{BH}=-\Xi_d^{-1}~\int_{m_{initial}}^{m_{min}}~{{dm}\over {x^{d+2}(m)T^{d+4}(m)}}\,, 
\end{equation}
with $m_{initial}$ being the original BH mass. We recall from above that while $x(m)$ is double valued it is well-behaved 
and never vanishes. However, on the otherhand, we also know that $T(m)\to 0$ as $m\to m_{min}$. Thus for all $n$ 
(and any $d$), $t_{BH}$ will be driven to infinity due to the presence of a singular denominator in the integrand 
implying a stable relic. (Of course, an initial very massive BH will radiate down to a mass very close to $m_{min}$ quite 
quickly.) This must happen in any model that predicts $T\to 0$ at finite $x$ due to a corresponding singularity. For the 
commuting case, since $T\sim 1/x$, the integrand is never singular so that the BH lifetime remains finite.   

The lower panel in Fig.~\ref{fig6} summarizes this discussion where we see that as $m$ decreases from a large value the 
temperature increases, reaches a maximum value and then falls to zero at $m_{min}$ leaving a classically stable remnant. 

\begin{figure}[htbp]
\centerline{
\includegraphics[width=8.5cm,angle=90]{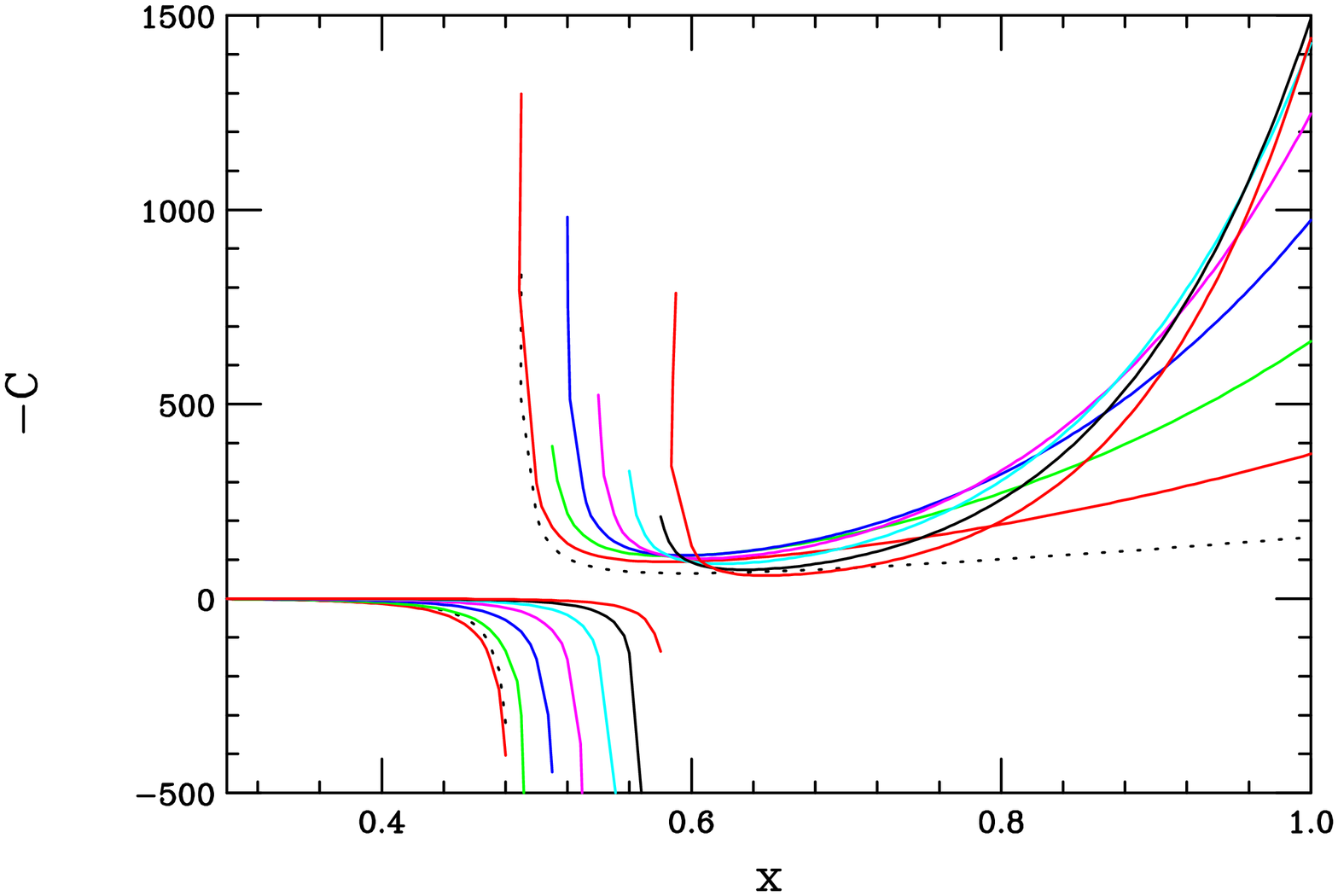}}
\vspace*{0.3cm}
\centerline{
\includegraphics[width=8.5cm,angle=90]{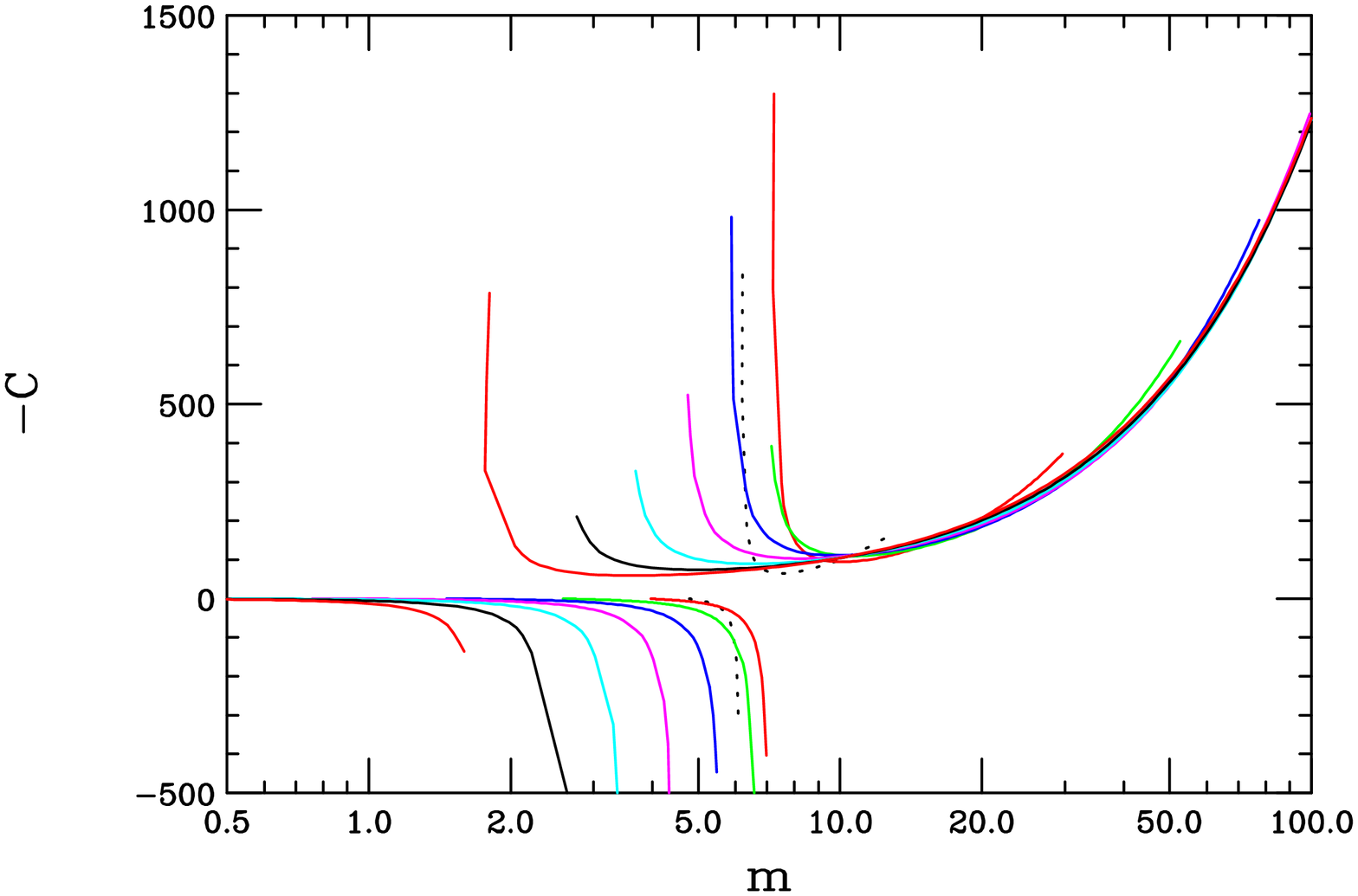}}
\vspace*{0.1cm}
\caption{Negative of the NC BH heat capacity/specific heat as a function of $x$(top panel) and $m$(bottom panel) 
for fixed $y=0.1$. The dotted curve 
corresponds to case of $n=0$ whereas, from bottom to top in the middle of the figure, the solid lines 
correspond to $n$=1 to 7. In the bottom panel, the parameter region with negative temperatures has been removed.}
\label{fig8}
\end{figure}

This unusual temperature behavior in the NC case 
can also be studied more fully by examining the BH heat capacity/specific heat, $C$. In the commutative 
case, $C$ is always finite (away from $x=0$) and negative since the BH gets hotter as it loses mass. Let we define  
\begin{equation}
C={{\partial m}\over {\partial T}}={{\partial m}\over {\partial x}}\Bigg({{\partial T}\over {\partial x}}\Bigg)^{-1}\,,
\end{equation}
which we can explicitly write as 
\begin{equation}
C=-4\pi c_n x^{n+2}F_n(q)^{-1}\Bigg[{{1-{{2H_n(q)}\over {(n+1)}}\over {1-{{2H_n(q)}\over{(n+1)}}+{{4qH_n(q)}\over {(n+1)}}
\Big({(n+3)\over {2q}}-{H_n(q)\over {q}}-1\Big)}}}\Bigg]\,,
\end{equation}
where for convenience we have defined the set of auxiliary functions 
\begin{equation}
H_n(q)={{q^{(n+3)/2}e^{-q}}\over {F_n(q)\Gamma({(n+3)\over {2}})}}\,.
\end{equation}
Using this expression, Fig.~\ref{fig8} shows the results for the NC 
BH heat capacity as a function of both $x$ and $m$ for our standard choice of $y=0.1$. We first see that $C$ remains 
negative at large $x$ and asymptotes to its commutative value, $C=-4\pi c_n x^{n+2}$, as it should. 
Further we note that at $x=x_{min}$ (or at $m=m_{min}$) we find that 
$C \to 0$ as was expected. BH with mass $m_{min}$ are no longer capable of mass loss since they have zero temperature. 
Between these two regimes the behavior of $C$ is quite interesting. Consider $C$ as a function of large and decreasing 
$x$. $C$ at first decreases in magnitude as it does in the commutative case. However as we know from above, for some 
$n$-dependent $x$ value, $T$ reaches a maximum and then decreases. This implies that the magnitude of $C$ then increases 
and becomes 
singular for this particular $x$ value. For lower $x$ the sign of $C$ changes as now ${{\partial T}\over {\partial x}}>0$ 
and then approaches zero at $m_{min}$. This is even more obvious when we consider $C$ as a function 
of $m$. For large $m$ the BH 
radiates as in the commutative case as $C$ decreases in magnitude as $m$ is reduced. However, at some point the NC effects 
turn on and $-C$ begins to increase, becoming singular where  ${{\partial T}\over {\partial m}}=0$, the location 
of temperature the maximum. For smaller masses, as 
the BH radiates and gets lighter the temperature decreases so that we are now in a region of positive heat capacity. As $m$ 
decreases further to $m_{min}$, $C$ becomes zero for the remnant.

\begin{figure}[htbp]
\centerline{
\includegraphics[width=8.5cm,angle=90]{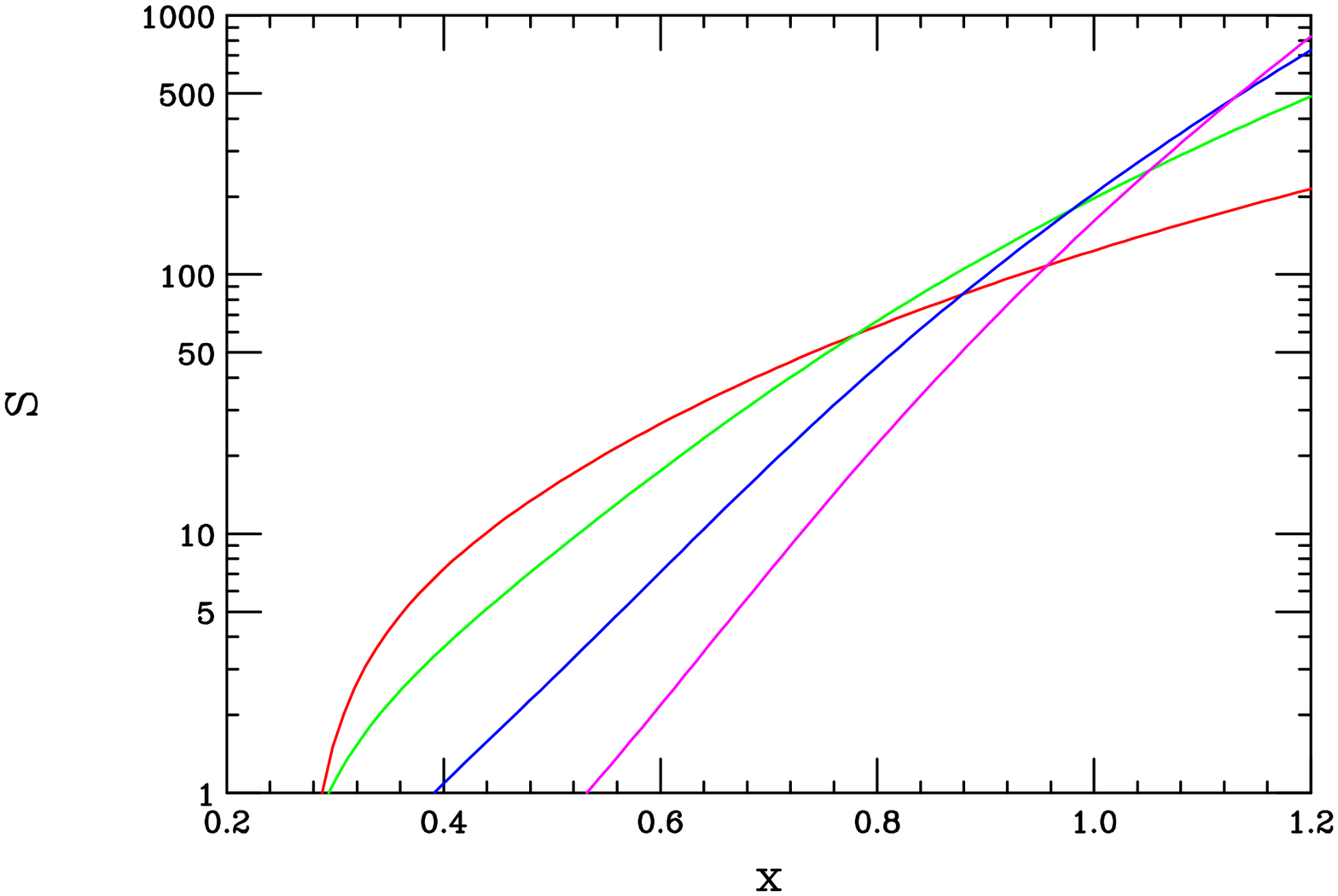}}
\vspace*{0.3cm}
\centerline{
\includegraphics[width=8.5cm,angle=90]{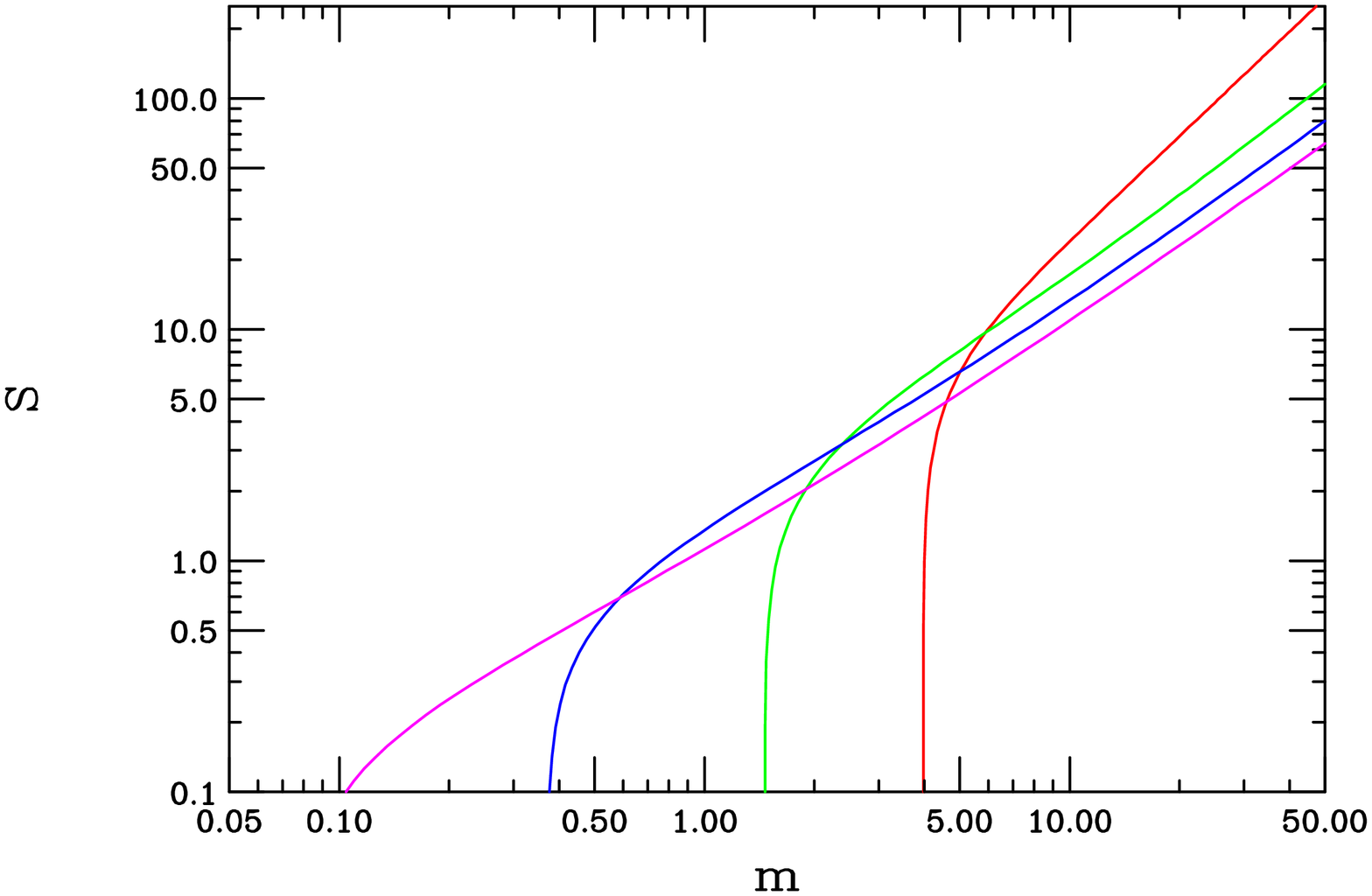}}
\vspace*{0.1cm}
\caption{The NC BH entropy as a function of $x$(top panel) and $m$(bottom panel) for fixed $y=0.1$. From left to 
right (right to left) at the bottom of the top(bottom) panel the the curves are for $n=1$, 3, 5 and 7, respectively.}
\label{fig9}
\end{figure}

Our next step in examining the thermodynamics of NC BH is to consider the value of the entropy which is defined via 
\begin{equation}
S=\int ~dx ~T^{-1}{{\partial m}\over {\partial x}}=4\pi c_n \int_{x_{min}}^x ~dv~{{v^{n+1}}\over {F_n(v^2/4y^2)}}\,,
\end{equation}
where here we have made the natural choice that the entropy vanish at $x_{min}$ where the BH mass $m$ is minimized 
for fixed values of $y$ 
and $n$. In the commutative limit where $\theta,y \to 0$ then $x_{min}\to 0$ and we recover the usual lower limit of 
integration. Fig.~\ref{fig9} shows the values of the entropy for various $n$ as a functions of $x$ or $m$ for our 
canonical choice of $y=0.1$. We again observe that the commutative power law behavior, $S=4\pi c_n {{x^{n+2}}\over {(n+2)}}$, 
is recovered in the large $x \gsim 1$ limit as expected. It is more interesting to consider $S$ as a function of $m$; as we 
see from the Figure, while the entropy scales as $\sim m^{(n+2)/(n+1)}$ for large $m$ it rapidly falls below this scaling 
law to zero as $m$ approaches $m_{min}$ from above.  

\begin{figure}[htbp]
\centerline{
\includegraphics[width=8.5cm,angle=90]{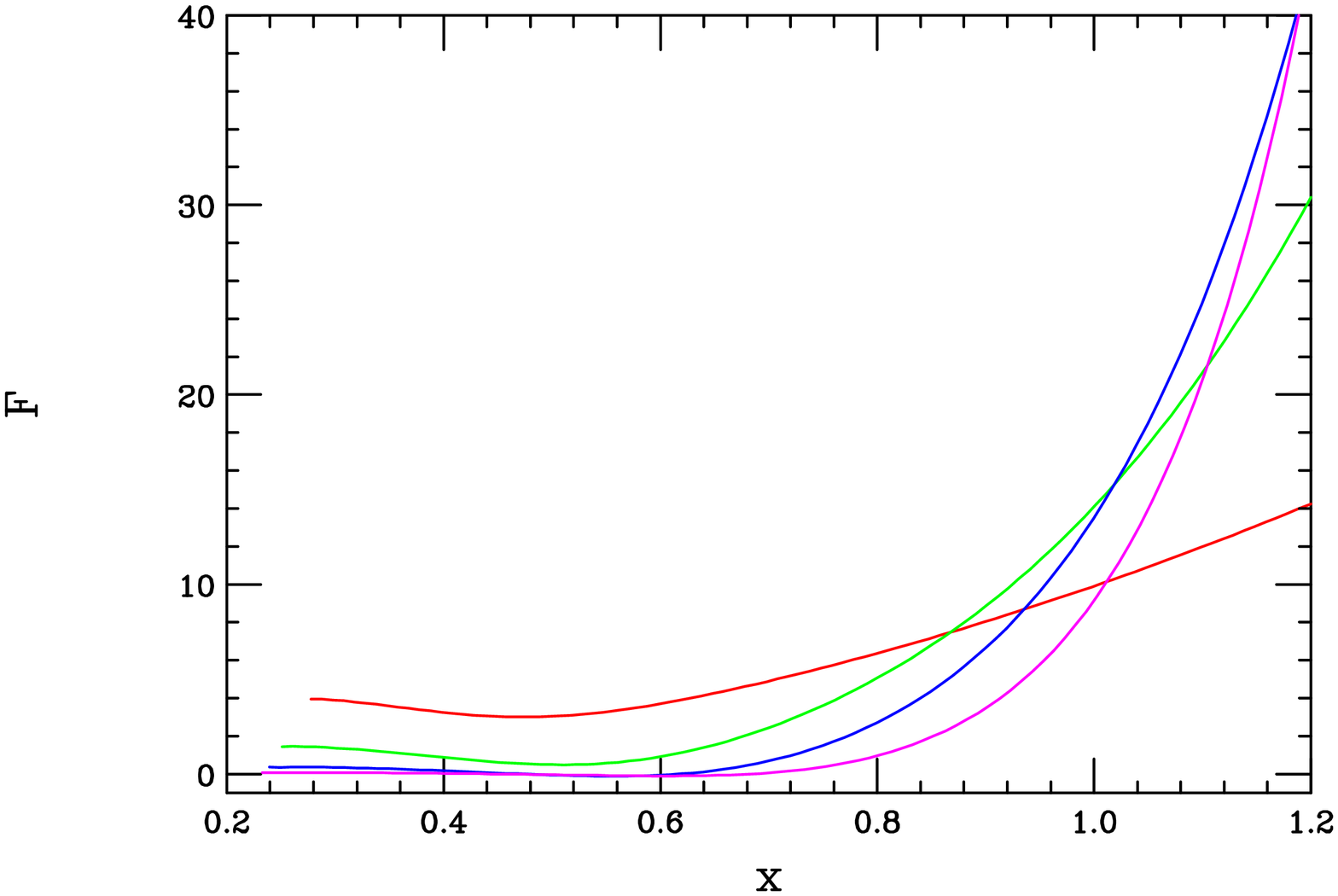}}
\vspace*{0.3cm}
\centerline{
\includegraphics[width=8.5cm,angle=90]{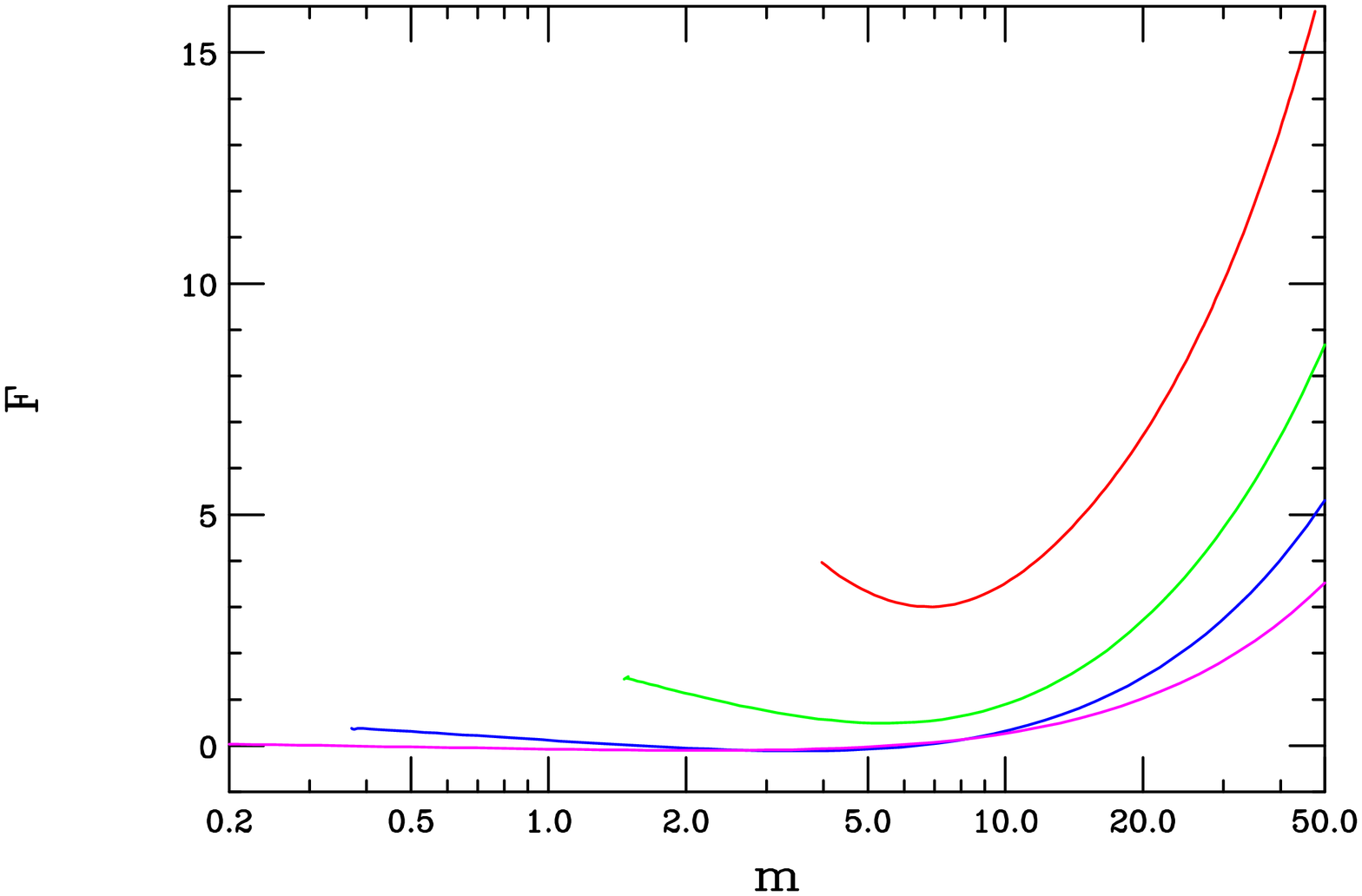}}
\vspace*{0.1cm}
\caption{The NC BH free energy as a function of $x$(top panel) and $m$(bottom panel) for $y=0.1$. From top to bottom 
on the left-hand side the curves correspond to $n=1$, 3, 5 and 7, respectively}
\label{fig10}
\end{figure}

Finally, we also examine the free energy of the NC BH which is simply given by combining the above thermodynamic quantities: 
\begin{equation}
F=m-TS\,.
\end{equation}
As seen in Fig.~\ref{fig10}, it too returns to its commutative value, $F=m/(n+2)$, in the limit of large $x\gsim 1$ or large 
$m$. However, for $x=x_{min}$ it is clear that $F=m$ for all values of $y$ and $n$ since both $T$ and (by definition) 
$S$ vanish at this point. 
Immediately above $x=x_{min}$ (or $m_{min}$), 
$F$ decreases slightly, until it matches onto the asymptotic $\sim x^{n+1}$ behavior. For 
some parameter values $F$ can even become negative in this intermediate mass regime. Generally we see the unusual behavior 
that $m=m_{min}$ is {\it not} the location of a minimum in $F$ as either a function of $x$ or $m$. This minimum 
lies near the values of $x$ and $m$ where $T$ is maximized. 

\section{Discussion and Conclusions}

Space-time noncommutativity and extra dimensions are both well-motivated ideas within the string theory context and it is 
natural for them to make their appearance felt as one approaches the fundamental scale. In more than four dimensions it is 
possible for this scale to be not far from $\sim$ TeV and thereby address the gauge hierarchy problem. If this mass 
parameter is indeed this low it is likely that black holes will be produced at the LHC in sufficient quantities that their 
properties will be well measured. The occurrence of NC also at a similar scale could lead to significant modifications in the 
anticipated properties of these BH. Since a complete NC theory of gravity does not yet exist it becomes necessary to model 
the NC effects within the commutative General Relativity framework. 

Nicolini, Smailagic and Spallucci presented a physically motivated model of this kind in 4-d, where the essential aspects 
were the Gaussian smearing of matter distributions on the NC scale and the continued applicability of the EH action. 
They then went on to examine NC effects on BH physics. In this paper we 
extended this NC BH analysis in several ways: ($i$) we generalized the NSS study to the case of extra dimensions with a 
fundamental scale in the TeV range so that the associated BH can be produced with large cross sections and studied in detail 
at the LHC. While much of the BH behavior observed in extra dimensions was similar to that obtained in 4-d, some significant 
modifications to the previously obtained 4-d results were observed. However, there appears to be an overall dominance of 
NC effects over those that arise due to the existence of the extra dimensions. ($ii$) We demonstrated 
that the essential physics induced by NC smearing is not particularly sensitive to the nature of the smearing 
function. In particular we explicitly showed that Gaussian and Lorentzian smearing lead to essentially the same behavior 
for the expected modifications of the BH mass-radius relationship due to NC effects. ($iii$) We extended the NSS analysis 
to include several other thermodynamic quantities which are of interest in the study of NC BH such as their entropy, heat 
capacity and free energy. 

Perhaps the most important qualitative influence of NC on BH physics was already observed in 4-d by NSS, \ie, 
the existence of a 
classically stable remnant whose mass and radius are completely fixed by the NC scale and the number of dimensions. Within 
the framework of extra dimensions, if the fundamental scale is not too large then BH and their remnants 
will be copiously produced at the LHC and 
studied in detail. The observation of NC effects in the properties of these BH can open a new window on the fundamental theory 
of gravity and space-time.

\noindent{\Large\bf Acknowledgments}

The author would like to thank J.L. Hewett and B. Lillie for discussions related to this work.

%
\def\MPL #1 #2 #3 {Mod. Phys. Lett. {\bf#1},\ #2 (#3)}
\def\NPB #1 #2 #3 {Nucl. Phys. {\bf#1},\ #2 (#3)}
\def\PLB #1 #2 #3 {Phys. Lett. {\bf#1},\ #2 (#3)}
\def\PR #1 #2 #3 {Phys. Rep. {\bf#1},\ #2 (#3)}
\def\PRD #1 #2 #3 {Phys. Rev. {\bf#1},\ #2 (#3)}
\def\PRL #1 #2 #3 {Phys. Rev. Lett. {\bf#1},\ #2 (#3)}
\def\RMP #1 #2 #3 {Rev. Mod. Phys. {\bf#1},\ #2 (#3)}
\def\NIM #1 #2 #3 {Nuc. Inst. Meth. {\bf#1},\ #2 (#3)}
\def\ZPC #1 #2 #3 {Z. Phys. {\bf#1},\ #2 (#3)}
\def\EJPC #1 #2 #3 {E. Phys. J. {\bf#1},\ #2 (#3)}
\def\IJMP #1 #2 #3 {Int. J. Mod. Phys. {\bf#1},\ #2 (#3)}
\def\JHEP #1 #2 #3 {J. High En. Phys. {\bf#1},\ #2 (#3)}

\end{document}